\documentclass[10pt,letterpaper]{article}
\usepackage[top=0.85in,left=2.75in,footskip=0.75in]{geometry}

\usepackage{adjustbox}
\usepackage{multicol}
\usepackage{multirow}
\usepackage[utf8]{inputenc}
\usepackage[english]{babel}
\usepackage{graphicx}
\usepackage{gensymb}
\usepackage{enumerate}
\usepackage[shortlabels]{enumitem}
\usepackage{float}
\usepackage{amsmath}
\usepackage{tabularx}
\usepackage{booktabs}
\usepackage{soul}
\usepackage{subcaption}
\usepackage{caption}      
\usepackage{amsmath,amssymb}

\usepackage{changepage}

\usepackage{textcomp,marvosym}

\usepackage{cite}

\usepackage{nameref,hyperref}

\usepackage[right]{lineno}

\usepackage[nopatch=eqnum]{microtype}
\DisableLigatures[f]{encoding = *, family = * }

\usepackage[table]{xcolor}

\usepackage{array}

\newcolumntype{+}{!{\vrule width 2pt}}

\newlength\savedwidth

\raggedright
\usepackage[aboveskip=1pt,labelfont=bf,labelsep=period,justification=raggedright,singlelinecheck=off]{caption}

\makeatletter
\renewcommand{\@biblabel}[1]{\quad#1.}
\makeatother

\newcolumntype{Y}{>{\centering\arraybackslash}X}
\newcolumntype{L}{>{\raggedleft\arraybackslash}p}
\newcolumntype{B}{>{\raggedright\arraybackslash}X}
\newcolumntype{C}{>{\centering\arraybackslash}m}

\usepackage{lastpage,fancyhdr,graphicx}
\usepackage{epstopdf}
\fancyheadoffset[L]{2.25in}
\fancyfootoffset[L]{2.25in}
\definecolor{emred}{RGB}{250, 15, 15}
\definecolor{emgreen}{RGB}{0, 180, 0}
\definecolor{emblue}{RGB}{34, 61, 240}

\begin{document}
\vspace*{0.2in}

\begin{flushleft}
{\Large
\textbf\newline{SENSOR: An ML-Enhanced Online Annotation Tool to Uncover Privacy Concerns from User Reviews in Social-Media Applications} 
}
\newline
Labiba Farah\textsuperscript{1*\textparagraph}
Mohammad Ridwan Kabir\textsuperscript{2\textparagraph},
Shohel Ahmed\textsuperscript{1},
MD Mohaymen Ul Anam\textsuperscript{1},
Md. Sakibul Islam\textsuperscript{1}
\\
\bigskip
\textbf{1} Software Engineering Lab (SEL), Department of Computer Science and Engineering, Islamic University of Technology (IUT), Boardbazar, Gazipur - 1704, Bangladesh.\\
\textbf{2} Systems and Software Lab (SSL), Department of Computer Science and Engineering, Islamic University of Technology (IUT), Boardbazar, Gazipur - 1704, Bangladesh.\\

\bigskip

* labibafarah@iut-dhaka.edu (LF)\\
\textparagraph These authors contributed equally to the work.

\end{flushleft}

\section*{Abstract}
The widespread use of social media applications has raised significant privacy concerns, often highlighted in user reviews. These reviews also provide developers with valuable insights into improving apps by addressing issues and introducing better features. However, the sheer volume and nuanced nature of reviews make manual identification and prioritization of privacy-related concerns challenging for developers. Previous studies have developed software utilities to automatically classify user reviews as --- \textit{privacy-relevant}, \textit{privacy-irrelevant}, \textit{bug reports}, \textit{feature requests}, etc., using machine learning. Notably, there is a lack of focus on classifying reviews specifically as --- \textit{privacy-related feature requests}, \textit{privacy-related bug reports}, or \textit{privacy-irrelevant}. This paper introduces SENtinel SORt (SENSOR), an automated online annotation tool designed to help developers annotate and classify user reviews into these categories. For automating the annotation of such reviews, this paper introduces the annotation model, \textbf{GRACE} (\textbf{GR}U-based \textbf{A}ttention with \textbf{C}BOW \textbf{E}mbedding), using Gated Recurrent Units (GRU) with Continuous Bag of Words (CBOW) and Attention mechanism. Approximately 16000 user reviews from seven popular social media apps on \textit{Google Play Store}, including \textit{Instagram}, \textit{Facebook}, \textit{WhatsApp}, \textit{Snapchat}, \textit{X (formerly Twitter)}, \textit{Facebook Lite}, and \textit{Line} were analyzed. Two annotators manually labeled the reviews using SENSOR, achieving a Cohen’s Kappa value of 0.87, ensuring a labeled dataset with high inter-annotator agreement for training machine learning models. Among the models tested, GRACE demonstrated the best performance (macro F1-score: 0.9434, macro ROC-AUC: 0.9934, and accuracy: 95.10\%) despite class imbalance. SENSOR demonstrates significant potential to assist developers with extracting and addressing privacy-related feature requests or bug reports from user reviews, enhancing user privacy and trust. Future research will focus on analyzing user reviews from a broader range of applications across various categories on \textit{Play Store}, improving classifier versatility and performance, and developing tool features to identify recurring privacy-related themes from user reviews for efficient feature integration or bug-fixing.

\nolinenumbers

\section*{Introduction}
    The pervasive use of smartphones has accelerated the development of smartphone applications catering to diverse needs from social media and productivity to health and finance, which can be downloaded from online app stores such as Google Play Store, Apple App Store, etc. \cite{alismail2023evaluating,ebrahimi2022unsupervised,nema2022analyzing,tao2020identifying}. Users generally communicate their experience and critical feedback, especially regarding privacy concerns, lack of features, bugs, technical issues, etc., in the form of user reviews on app stores, providing developers valuable insights into improving apps by addressing reported issues and introducing better features \cite{alismail2023evaluating,hatamian2019revealing,nguyen2019short,maalej2016automatic}. 
    With smartphone applications becoming more embedded in everyday life \cite{ebrahimi2022unsupervised}, users are becoming more aware of data security and privacy risks than ever before \cite{degirmenci2020mobile,zhang2025mining}. Consequently, the volume and diversity of user-generated app reviews have grown significantly in recent times \cite{hatamian2019revealing,nema2022analyzing}. Although only 0.5\% of these reviews are related to security and privacy concerns \cite{mukherjee2020empirical}, they are invaluable to developers, as they help maintain privacy standards and reinforce user trust, unlike feedback on non-privacy-related issues \cite{nema2022analyzing,degirmenci2020mobile,hatamian2019revealing,ebrahimi2022unsupervised}. 

    Privacy-related reviews are often domain-specific, complex, unstructured, and nuanced \cite{nema2022analyzing,ebrahimi2022unsupervised}. For instance, some users may express concerns over apps accessing sensitive information, such as live location or banking details \cite{ebrahimi2022unsupervised,degirmenci2020mobile,zhang2025mining}. Others, particularly on social media platforms, may highlight issues like dissented surveillance, secondary usage of users' personal information, hacked accounts, lack of privacy features, etc. \cite{degirmenci2020mobile,alismail2023evaluating}. Consequently, manual analysis is challenging for developers as it is both time-consuming and inconsistent, often missing critical privacy issues and reducing the effectiveness of responses \cite{ebrahimi2022unsupervised,hatamian2019revealing}. Previous studies \cite{nema2022analyzing,ebrahimi2022unsupervised,zhang2025mining}, while effective in classifying general concerns, fall short of addressing privacy-specific annotations, particularly for privacy-related bugs and feature requests. Existing tools largely focus on general sentiment analysis and broad categorization, lacking a specialized focus on privacy \cite{besmer2020investigating,hatamian2019revealing,tao2020identifying}. 

    Literature \cite{maalej2016automatic} suggests a potential granular classification of privacy-related user reviews in terms of \textit{feature requests} and \textit{bug reports}. Despite significant advances in automated annotation and classification of user reviews, few studies specifically address the unique nature of privacy concerns within these reviews \cite{maalej2016automatic,harkous2018polisis}. This highlights the scope for developing an automated annotation tool to classify user reviews with more granularity in terms of privacy as --- \textit{privacy-related feature requests}, \textit{privacy-related bug reports}, and \textit{irrelevant}. In \textit{privacy-related feature requests}, users may ask for clarification for missing privacy-related features, e.g., ``Why remove the edit button? It’s our post, and we should have the freedom to edit or delete it. Give us back the choice!'', or they could provide suggestions for additional features that they perceive to play an important role in upholding their privacy, e.g., ``Add who saw your note that would make the app a little better!'' In \textit{privacy-related bug reports}, on the other hand, users tend to report privacy-related flaws in the functionality of various app features, e.g., ``After the new IOS update they keep logging me out after I chose for the app not to track me'' \cite{alismail2023evaluating,maalej2016automatic}. Reviews where none of the above are listed, e.g., ``The app keeps crashing and kicking me off!! It's so annoying!'', or generic frustrations related to privacy policies are reported, e.g., ``A garbage app that is not capable of anything but tracking you and your interests so they can sell your info'', may be categorized as \textit{privacy-irrelevant}.

    In this paper, we introduce \textbf{SEN}tinel \textbf{SOR}t (SENSOR), a machine learning-enhanced online annotation tool designed to annotate and classify privacy-related concerns in user-generated reviews of smartphone applications into the aforementioned categories. The main objective of our tool is to empower app developers to stay ahead of privacy concerns by identifying and resolving user-reported issues quickly and efficiently. SENSOR features two unique user roles --- \textit{Developer} and \textit{Annotator}. Developers can upload CSV files containing user reviews or retrieve reviews directly from the Google Play Store by providing the \textit{app-ID}, \textit{start-date}, and \textit{end-date}. After uploading the reviews, developers can assign two human annotators for manual annotation or utilize the pre-trained machine-learning model integrated into SENSOR for automated annotation.

    In manual annotation, annotators are restricted to accessing only the files assigned to them by the developer. They can categorize each review into one of the three predefined categories --- \textit{privacy-related feature requests}, \textit{privacy-related bug reports}, or \textit{privacy-irrelevant}. Developers can monitor the annotation progress for uploaded files, including the number of reviews annotated by each of the annotators. Once an annotator completes annotating all reviews in a file, that file becomes unavailable for further annotation by that annotator. A file is deemed fully annotated when both annotators have completed annotating all reviews and can be downloaded in \texttt{.csv} format. SENSOR automatically evaluates annotation quality using Cohen’s Kappa measure of inter-rater agreement \cite{rau2021evaluation, wang2022coolted}. Based on this assessment, developers can either accept the annotations or reassign the file to a new pair of annotators. This role-based workflow facilitates efficient collaboration between developers and annotators, enabling a systematic and scalable analysis of user reviews.
    
    To train the machine learning model for automated annotation, approximately 16,000 user reviews were collected from seven popular social media applications on the Google Play Store, including Instagram, Facebook, WhatsApp, Snapchat, X (formerly Twitter), Facebook Lite, and Line, based on their global impact, extensive feature sets, and the diversity of associated privacy challenges \cite{jain2021online, page2022social, miller2022don}. The first two authors independently annotated the reviews using the SENSOR tool, yielding a Cohen’s Kappa value of 0.87, reflecting a high level of agreement between the annotators \cite{rau2021evaluation}. The discrepancies in labeling were resolved through discussion between the two authors to reach a consensus. Following pre-processing, the dataset was partitioned into training, test, and validation sets according to the 80-10-10 rule, resulting in 12,756, 1,594, and 1,595 samples, respectively. The training set was augmented to 121,374 samples using popular Natural Language Processing (NLP) techniques \cite{joshi2023text} to train a wide range of classical machine learning models with various text representation methods, including TF-IDF \cite{qureshi2021performance} and word2vec \cite{church2017word2vec}. We also evaluated several deep-learning models, including our proposed deep-learning framework, \textbf{GRACE} (\textbf{GR}U-based \textbf{A}ttention with \textbf{C}BOW \textbf{E}mbedding), that leverages Gated Recurrent Units (GRU) \cite{dey2017gate} integrated with Continuous Bag of Words (CBOW) \cite{xia2023continuous} and Attention Mechanism \cite{brauwers2021general}. Among the models evaluated, GRACE achieved the most promising results, having a macro-averaged 94.83\% precision, 93.89\% recall, F1-score of 94.34\%, and a Receiver Operating Characteristic Area Under the Curve (ROC-AUC) value of 99.28\% on the test dataset. These results demonstrate the superior classification performance of the model, highlighting its effectiveness in the automated annotation task. 
    
    In summary, the contributions of this research are as follows ---
    \begin{enumerate}
        \item The research introduces SENSOR (available \href{https://sensor-1.onrender.com/}{here}), a machine learning-enhanced annotation tool that classifies privacy-related concerns in user reviews of social media apps into --- \textit{privacy-related feature requests}, \textit{privacy-related bug reports}, and \textit{privacy-irrelevant}, using either manual annotation or the pre-trained, tool-integrated machine learning model, addressing a gap in existing annotation tools. 
        
        \item The research introduces a manually annotated dataset of 16,000 user reviews focused on \textit{privacy-related feature requests} and \textit{privacy-related bug reports}, with a Cohen's Kappa coefficient of 0.87. This dataset (available \href{https://doi.org/10.5281/zenodo.15852840}{here}) was used to train and evaluate the tool-integrated machine learning model, offering a valuable resource for future privacy-focused research on user reviews.

        \item The study develops an automated annotation model, \textbf{GRACE} (\textbf{GR}U-based \textbf{A}ttention with \textbf{C}BOW \textbf{E}mbedding), using Gated Recurrent Units (GRU) \cite{dey2017gate} with Continuous Bag of Words (CBOW) \cite{xia2023continuous} and Attention Mechanism \cite{brauwers2021general}, achieving macro-averaged precision, recall, and F1-scores above 93\%, demonstrating its effectiveness in classifying privacy concerns in user reviews.
    \end{enumerate}
    
    The remainder of the paper is organized as follows. We begin by reviewing relevant literature, focusing on studies that address privacy-related concerns in user feedback and exploring the development of user review mining and annotation tools, which together form the foundation for our contribution. This is followed by an overview of the working mechanism and the core functionalities of the proposed SENSOR tool. We then describe the research methodology, outlining the procedures for scraping, filtering, and labeling user reviews, as well as the data pre-processing and augmentation strategies applied. The classification pipeline is detailed next, encompassing the development and training of the proposed GRACE framework as well as classical machine learning and state-of-the-art deep learning frameworks, followed by an insight into the models' performance evaluation protocol. The results section presents a comprehensive analysis of classifier performance and inference efficiency. Finally, we conclude with a discussion of key findings, broader implications for privacy engineering, and potential avenues for future research.

\section*{Related Work}
    The rapid growth in the development of social media applications has led to an exponential increase in user-generated reviews. Several studies have explored the intersection of privacy concerns and user reviews in mobile applications, utilizing machine learning techniques to classify and address these issues. This section presents a thorough comparison of SENSOR with existing literature, emphasizing its distinct contributions and demonstrating how it addresses gaps within the current body of research.
    
    \subsection*{Focus on Privacy-Related Concerns}        
        Zhang et al. used datasets from 11 apps across four categories to classify reviews into binary categories --- \textit{privacy related} and \textit{non-privacy related}, followed by detection of privacy concern topics using various topic modeling algorithms. For review classification, they employed models like Gradient Boosting and BERT, achieving an F1-score of 93.77\% without further distinguishing between feature requests and bug reports \cite{zhang2023mining, zhang2022detecting}. Mukherjee et al. analyzed 2.2 million user reviews from 539 top free apps in the Google Play Store to understand how much users are concerned about security and privacy in mobile apps and whether their concerns align with the actual behavior of the app after categorizing the reviews as \textit{privacy related} or \textit{non-privacy related}. They sampled about 4122 potential security and privacy-related reviews using a list of privacy-related keywords, such as privacy, security, access, permission, steal, secure, etc., and about 1878 reviews without any such keywords. Their final training set, after manual annotation and text processing, contained 936 \textit{privacy related} and 5046 \textit{non-privacy related} reviews. They trained \textit{four} classical machine learning models, including Support Vector Machine (SVM), which achieved the best results \cite{mukherjee2020empirical}. Besmer et al. analyzed user perceptions of mobile app privacy by classifying reviews as \textit{privacy-related}, those that mention privacy concerns or praise, and \textit{non-privacy-related}, those that do not discuss privacy at all, using a logistic regression classifier. Their dataset includes over five million reviews from the Amazon App Store. However, they manually annotated a small subset of 2372 user reviews (398 \textit{privacy-related} and 1974 \textit{non-privacy-related}) reviews for training the classifier. They also use AFINN sentiment analysis to compare sentiment scores between the two groups, with privacy-related reviews having lower star ratings and more negative sentiment \cite{besmer2020investigating}. Nema et al. presented a large-scale analysis of about 287 million reviews from 2 million apps across 29 categories in the Google Play Store, leveraging NLP techniques, including deep learning models like Bidirectional Encoder Representations from Transformers (BERT) \cite{devlin2018bert} and Universal Sentence Encoder (USE) \cite{cer2018universal}, to classify, cluster, and summarize privacy-related reviews. Three annotators participated in manually annotating 11371 reviews (6688 \textit{privacy related} and 4683 \textit{non-privacy related}) for training and validation. Disagreements among the annotators were resolved through discussion. The authors used an ensemble of BERT and USE models to distinguish between \textit{privacy related} and \textit{non-privacy related} reviews with 98\% precision, 87\% recall, an F1-score of 92.49\%, and an ROC-AUC value of 98.18\%. They also applied K-means clustering to group similar privacy-related reviews into themes like excessive permissions, tracking, and selling data, etc., and used summarization metrics to extract representative reviews for each cluster \cite{nema2022analyzing}. Nguyen et al. employed supervised machine learning techniques to classify reviews into \textit{security and privacy-related reviews} (SPR) and non-SPR. They used static code analysis to map SPR to app updates and evaluated the impact of SPR on app security and privacy \cite{nguyen2019short}.

        Mehrdad et al. proposed a deep learning-based model, DARCLSTM, for classifying user reviews in the Google Play Store into three categories --- \textit{feature requests}, \textit{bug reports}, and \textit{information giving}. They utilized 51,000 reviews from 10,842 apps across 32 categories for training and evaluating the model, achieving an F1-score of 95.3\% and an accuracy of 97.21\% \cite{dehkordiauser}. McIlroy et al. classified user reviews from mobile apps into three categories --- \textit{feature requests}, \textit{bug reports}, and \textit{privacy complaints}. They achieved the best performance using Pruned Sets with SVM with 66\% precision and 65\% recall \cite{mcilroy2016analyzing}. Combining text classification with metadata and NLP techniques, Maalej et al. classified app reviews into four categories --- \textit{feature requests}, \textit{bug reports}, \textit{user experiences}, and \textit{ratings} with precision and recall ranging between 88–92\% and 90-99\%, respectively. They utilized a dataset of 4400 manually labeled reviews for training and evaluating classical machine learning models like Naive Bayes, Decision Trees, and Max Entropy. They discovered that multiple binary classifiers outperformed a single multi-label classifier, as no single classifier worked best for all review types and data sources \cite{maalej2016automatic}. Kaur et al. proposed a hybrid deep learning framework, BERT-RCNN, to classify app reviews into specified categories of functional requirements such as --- \textit{feature requests}, \textit{bug reports}, and \textit{relevant} etc. The authors used five manually annotated benchmark datasets for app review classification, including reviews from Google Play and App Store, with their model achieving precision, recall, and F1-score values ranging between 81-96\%, 64-89\%, and 69-91\%, respectively. Their work highlighted the importance of contextual information and transfer learning in improving the classification accuracy of nuanced user reviews \cite{kaur2023bert}. Hatamian et al. analyzed user reviews of 200 apps across 10 categories on the Google Play Store and categorized them into predefined privacy threats, including \textit{phishing}, \textit{targeted ads}, \textit{spam}, \textit{unauthorized charges}, etc. They employed various classical machine learning models, NLP, and sentiment analysis techniques, and achieved overall precision, recall, and F-score values of 94.84\%, 91.30\%, and 92.79\%, respectively, using Logistic Regression \cite{hatamian2019revealing}. Reddy et al. use a modified logistic regression technique to classify movie reviews into \textit{positive}, \textit{negative}, and \textit{neutral} sentiments. Their model achieves an accuracy of 90\%, but it is limited to sentiment analysis and does not address privacy concerns \cite{reddy2024classification}. Qureshi et al., on a similar footing, evaluated traditional ML models like SVM, Random Forest, and Naive Bayes for sentiment analysis of app reviews \cite{qureshi2021performance}.
        
        While the studies mentioned above offer comprehensive frameworks for user review classification, which could be valuable to diverse stakeholders, they do not specifically target feature requests or bug reports related to privacy. In contrast, SENSOR specifically targets privacy-related concerns in user reviews, classifying them into three granular categories: \textit{privacy-related feature requests}, \textit{privacy-related bug reports}, and \textit{privacy-irrelevant} reviews. SENSOR achieves a macro-averaged F1-score of 94.34\% and an ROC-AUC of 99.28\%, while using the proposed deep-learning framework GRACE. This demonstrates the superiority of SENSOR's deep learning approach over traditional methods, making it more effective for large-scale review analysis. This focus aligns with the growing importance of data protection regulations like the General Data Protection Regulation (GDPR) and the California Consumer Privacy Act (CCPA), making SENSOR a timely and relevant tool for developers \cite{fakeyede2023navigating}.
        
    \subsection*{Focus on User Review Mining and Annotation Tool}
        A significant limitation of many existing tools is the lack of a structured workflow for collaboration between developers and annotators. Tao et al. introduce SRR-Miner, a tool designed for developers to analyze user reviews without any explicit support for collaborative workflows or role-based access. It uses a manually-created list of security-related keywords to extract relevant issues from mobile app reviews, including system vulnerabilities, privacy violations, and malicious behaviors, with an F1-score of 0.89. It then summarizes these reviews as $<$\textit{misbehavior-aspect-opinion}$>$ triples, identified employing semantic patterns. The triples capture the app's \textit{misbehavior}, the \textit{aspect} of the app being discussed, and the user's \textit{opinion}. The summary is visualized using a radar chart to help target users understand the frequency and nature of security issues. Although the tool does not focus on privacy-specific concerns, it aims to help users understand security issues in apps by analyzing user reviews, particularly focusing on negative and neutral sentiments \cite{tao2020identifying}.     

        Contrasting with SRR-Miner, Wang et al. developed CoolTeD as an online collaborative tool to help developers with labeling software requirements as --- \textit{functional} and \textit{non-functional}, following ISO 25010. It provides a structured approach to labeling with predefined categories for \textit{non-functional} requirements. The tool fosters teamwork by allowing multiple coders to work independently on the same dataset. To measure the accuracy of the labeling process, CoolTeD calculates Cohen’s Kappa coefficient to gauge the inter-rater agreement. It also supports confidence levels and rationale recording to improve the quality of labels. Additionally, it features a review system for senior coders to resolve disagreements and ensure consistency across labels while storing the annotated results in a simple JSON file. CoolTeD offers visualization options like pie charts to show the distribution of labeled data and allows users to track the labeling process over time through its web interface. Despite all these features, it does not specifically focus on privacy reviews and lacks automated annotation employing machine learning. Hatamian et al. introduced the Mobile App Reviews Summarization (MARS) tool to help developers understand privacy complaints and improve their apps' privacy practices \cite{hatamian2019revealing}.

        Gu et al. developed SUR-Miner to help developers understand user preferences by classifying app reviews as \textit{bug reports} and \textit{feature requests}, followed by extracting aspect-opinion pairs. It also uses visualizations, such as the Aspect Heat Map, to track sentiment changes over time \cite{gu2015parts}. In contrast, SURF by Di Sorbo et al. uses the User Reviews Model (URM) to classify reviews based on user intent and app topics, turning feedback into actionable tasks like bug fixes. Tested on 17 apps, SURF helps developers quickly prioritize software changes, saving time \cite{di2016would}. While both tools simplify review analysis, SUR-Miner focuses on sentiment and aspects, whereas SURF provides structured summaries for actionable tasks. Panichella et al. developed ARdoc, a tool that combines NLP, text analysis, and sentiment analysis to automate the classification of sentences into five categories --- \textit{feature requests}, \textit{problem discovery}, \textit{information seeking}, \textit{information giving}, and \textit{other}, extracting actionable insights to improve app-functionality through efficient analysis of large volumes of reviews. The tool achieves high classification accuracy, with precision, recall, and F-measure ranging between 84\% and 89\%. ARdoc is available both as a GUI tool and a Java API, making it versatile for different use cases \cite{panichella2016ardoc}. Iacob et al. designed the Mobile App Review Analyzer (MARA), a tool to help developers automatically gather \textit{feature requests} from user reviews of mobile apps. It analyzes feedback using predefined linguistic rules to spot feature requests, and in a study of over 136,000 reviews, it found that about 23.3\% contained suggestions for improvements like better support, regular updates, and more customization options. By using Latent Dirichlet Allocation (LDA), MARA also identifies recurring topics across these requests. In essence, MARA saves developers time by automatically extracting valuable insights from user feedback, making it easier to prioritize app enhancements \cite{iacob2013retrieving}. Fu et al. designed WisCom to extract user ratings and comments from the Google Play Store to analyze user feedback on mobile apps in three levels --- micro-level (individual reviews), meso-level (specific app-level reviews), and macro-level (overall market trends). The system was tested on a large dataset of over 13 million reviews to provide actionable insights for developers, end-users, and market operators to improve app quality and user experience \cite{fu2013people}.
        
        SENtinel SORt (SENSOR) offers a significant advancement over existing tools by providing an integrated, machine-learning-enhanced platform for annotating and classifying privacy-related concerns in user reviews, particularly addressing the gap in tools that focus on privacy-specific issues. Unlike tools such as SRR-Miner, which focuses primarily on security issues without supporting collaborative workflows, SENSOR introduces a structured, role-based system that allows both developers and annotators to collaborate efficiently. Additionally, the automated annotation feature using the integrated machine learning model makes it more suitable for large-scale analysis, a feature lacking in many existing tools like CoolTeD, MARS, and SUR-Miner, which do not incorporate privacy-specific annotations or fully automated approaches. Moreover, SENSOR facilitates real-time retrieval of reviews directly from the Google Play Store by providing the \textit{app-ID}, \textit{start-date}, and \textit{end-date}, a unique feature not present in the existing tools, making it a valuable addition to the developer community, especially as privacy regulations continue to evolve.
   
\section*{SENSOR Tool}
    The \textbf{SEN}tinel \textbf{SOR}t (SENSOR) tool is a machine learning-enhanced online annotation tool designed to annotate and classify privacy-related concerns in user-generated reviews of smartphone applications through two user roles --- \textit{Developer} and \textit{Annotator}. The core features and functionalities of SENSOR revolve around the themes --- \textit{user authentication and role management}, \textit{review acquisition and annotation workflow}, \textit{progress monitoring and feedback mechanism}. This section provides an in-depth overview of each of these themes.
    
    \begin{figure}[ht]
        \centering
        [FIGURE \ref{fig:user_auth}]
        \vspace{5mm}
        
        \includegraphics[width=.75\textwidth]{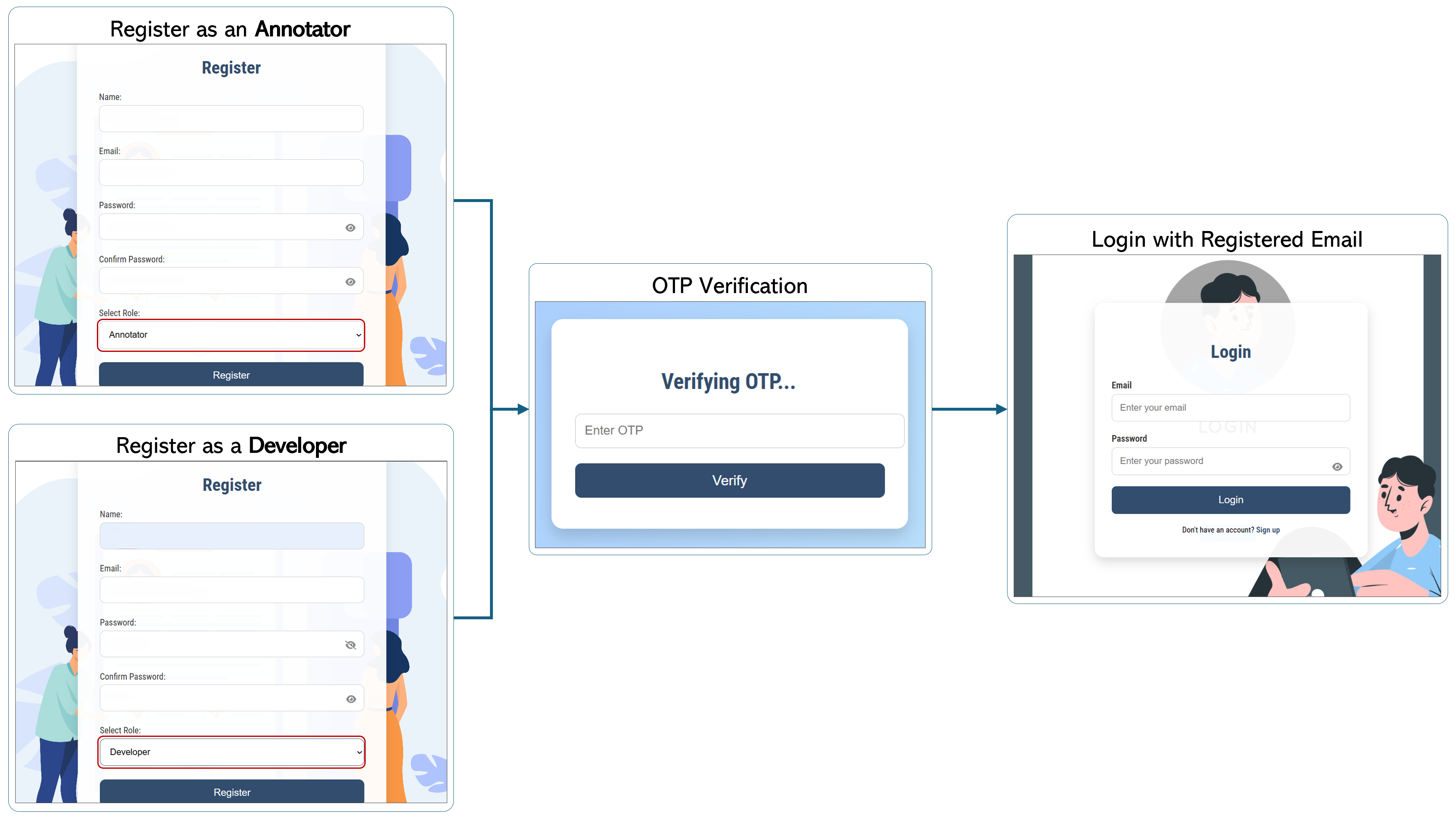}
        \caption{\textbf{User registration, authentication, and login workflow of SENSOR tool. Users can either register as a \textit{Developer} or an \textit{Annotator}, followed by an OTP verification step, can login to the system using the registered email and password.}}
        \label{fig:user_auth}
    \end{figure}
    
    \subsection*{User Authentication and Role Management}
        The platform incorporates a secure and robust user authentication framework built using Flask-Login and Flask-WTF \cite{lado2025flask}. During registration, users are required to complete a One-Time Password (OTP) verification step to ensure identity validation. Upon successful verification, a user session is initiated and maintained throughout the user’s engagement with the system. The process is summarized in \autoref{fig:user_auth}.
        
\begin{figure}[ht]
            \centering
            [FIGURE \ref{fig:developer_dash}] 
            \vspace{5mm}
            \includegraphics[width=\textwidth]{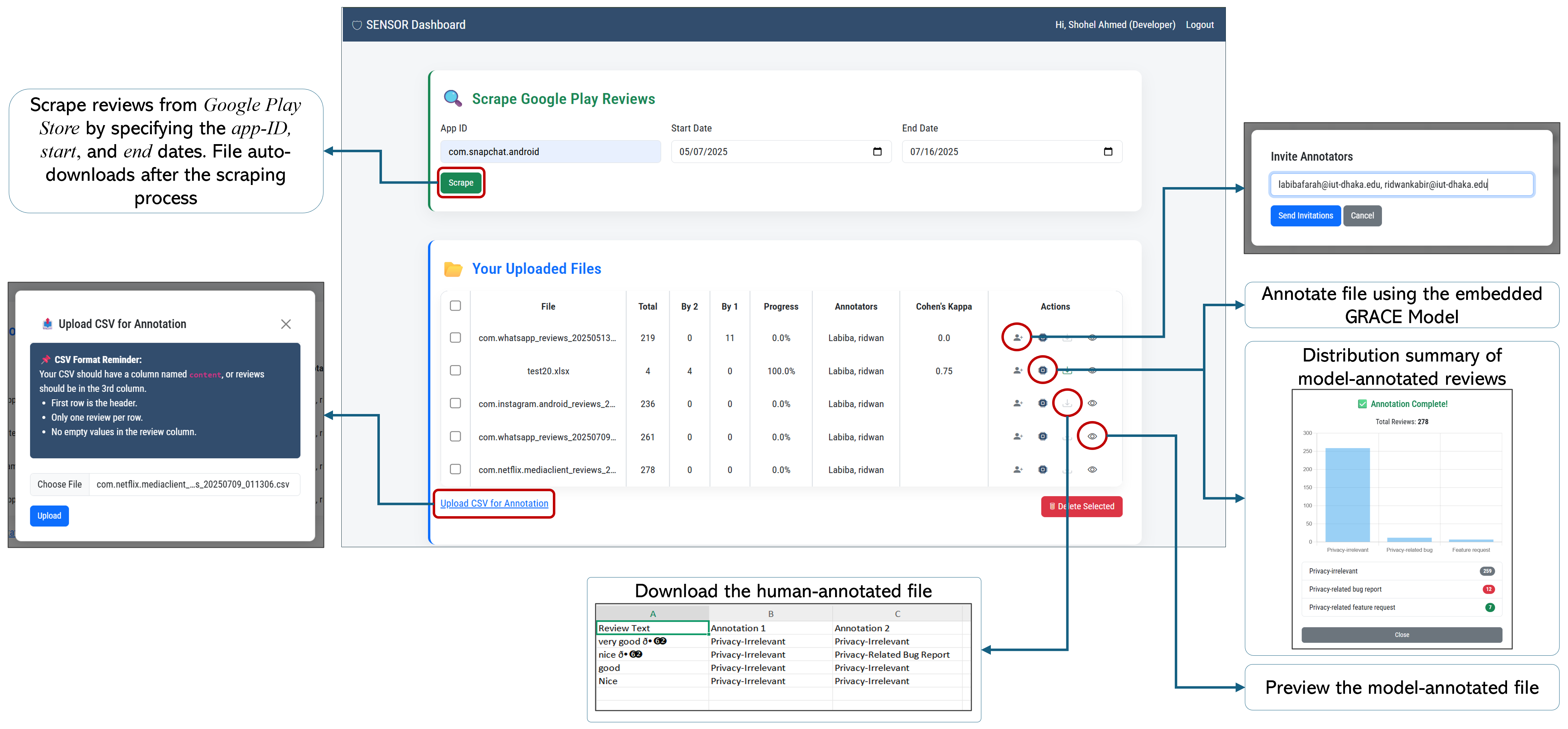}%
            \caption{\textbf{Developer dashboard of SENSOR tool. Developers can scrape Google Play reviews by specifying an \textit{app-ID} and \textit{date range}, upload CSVs for annotation, invite annotators, track progress and quality of human annotation, perform model-based (GRACE) annotation, download human-annotated files, and provide feedback on model-generated annotations.}}
            \label{fig:developer_dash}
        \end{figure}

        SENSOR features two types of users, \textit{Developers} and \textit{Annotators}, each with distinct responsibilities and access privileges, and based on the user role, two separate dashboards are provided, as shown in \autoref{fig:developer_dash} and \autoref{fig:annotator_dash}, respectively. Developers are granted administrative capabilities, including the ability to \textit{scrape user reviews}, \textit{manage CSV files}, \textit{assign human annotators}, \textit{monitor annotation progress}, and \textit{oversee the use of automated annotation models} (\autoref{fig:developer_dash}). Annotators, on the other hand, can label user reviews, track, and save their current progress. 
        
    \subsection*{Review Acquisition and Annotation Workflow}
        SENSOR offers a flexible and efficient review acquisition pipeline, enabling developers to collect user feedback from the Google Play Store using the \texttt{google-play-scraper} library \cite{hasanah2025play}. Developers can configure the parameters --- \textit{app ID} (a unique identifier for the target application) and a \textit{date range} (to filter reviews based on their publication dates) for the scraping process. The scraped reviews are then auto-downloaded as a CSV file, which the developers can upload to the online database of SENSOR manually, serving as input for either human or model annotation (\autoref{fig:developer_dash}). The system performs several pre-processing steps on the uploaded file, including format normalization, handling of missing values, and automatic standardization of column names to ensure consistency across datasets.

        
        \begin{figure}[ht]
            \centering
            [FIGURE \ref{fig:annotator_dash}] 
            \vspace{5mm}
            \includegraphics[width=\textwidth]{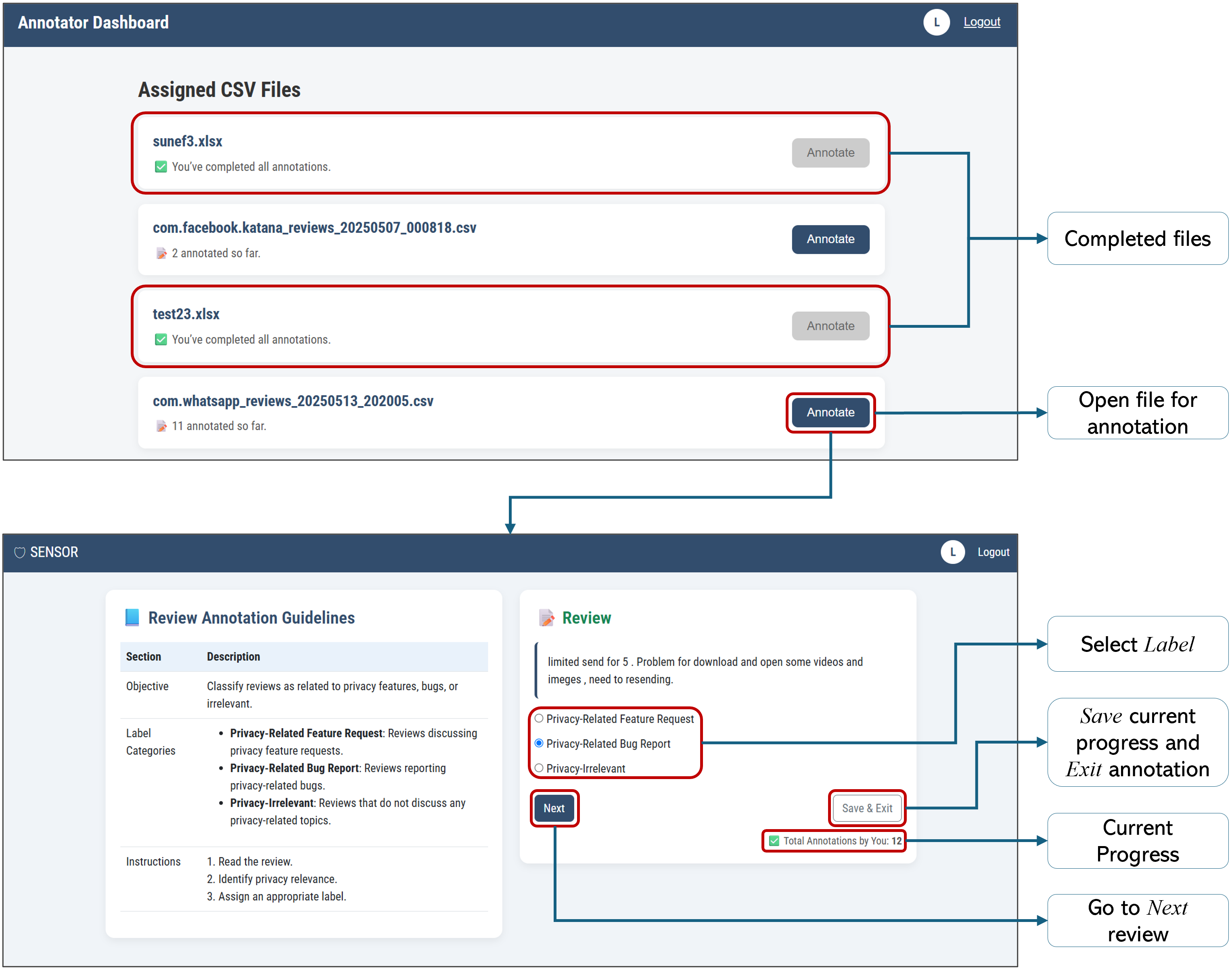}%
        
            \caption{\textbf{Annotator dashboard of Sensor tool. Annotators can access files they have been invited to and label reviews as \textit{privacy-related feature requests}, \textit{privacy-related bug reports}, or \textit{privacy-irrelevant} following the provided guidelines. They can also monitor and save their current progress.}}
            \label{fig:annotator_dash}
        \end{figure}
        
        Following data acquisition, SENSOR facilitates a structured and role-based annotation process. Developers can assign each uploaded dataset to two independent annotators through email invitation. Annotators can only access review files after receiving the email invitation from a developer. These invitations include direct links to the assigned files along with detailed instructions, ensuring a structured and guided annotation process. Annotators can then label individual reviews in a certain file based on predefined categories (\textit{privacy-related feature requests}, \textit{privacy-related bug reports}, or \textit{privacy-irrelevant}) (\autoref{fig:annotator_dash}). To further support efficiency, SENSOR integrates a model-driven annotation option powered by the GRACE framework (\autoref{fig:developer_dash}). SENSOR allows developers to download human-annotated and model-annotated files separately in \texttt{.csv} format, which they can manually analyze to make decisive actions. This dual approach enhances the scalability, efficiency, and accuracy of the annotation process, facilitating high-quality labeled datasets for downstream analysis or machine learning applications.

    \subsection*{Progress Monitoring and Feedback Mechanism}
        SENSOR incorporates comprehensive tracking features that enable developers to oversee the annotation process with clarity and precision. In the case of human annotation, SENSOR provides a real-time summary of annotation progress, including the number of reviews annotated by both assigned annotators and the percentage of reviews completed (a review is considered fully annotated when both assigned annotators independently annotate it). This overview enables developers to easily monitor workflow status and identify any bottlenecks in the annotation pipeline. To assess the consistency of human annotation, SENSOR calculates the Cohen’s Kappa score \cite{rau2021evaluation}, a widely used statistical measure that quantifies inter-rater agreement, in real time (\autoref{fig:developer_dash}). By doing so, the system helps ensure that both annotations reflect a shared understanding of the labeling criteria across different annotators. 

        \begin{figure}[ht]
            \centering
            [FIGURE \ref{fig:feedback_mechanism}]
            \vspace{5mm}
            \includegraphics[width=\textwidth]{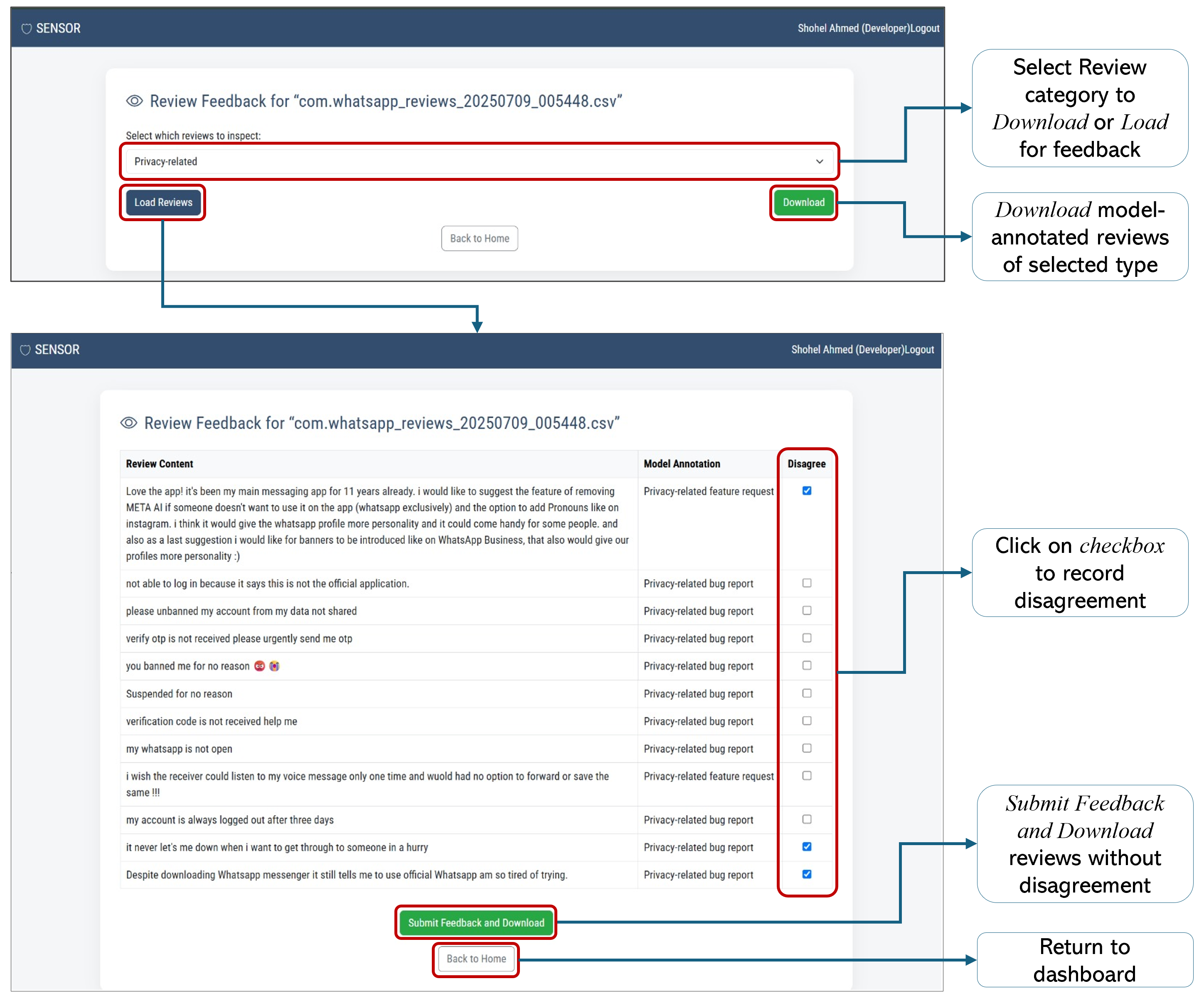}
            \caption{\textbf{Developer feedback module of SENSOR tool for model-annotated reviews. Developers can filter model-annotated reviews by category (\textit{privacy-related} or \textit{privacy-irrelevant}), inspect annotations, record disagreements, submit feedback to improve reliability, and download results.}}
            \label{fig:feedback_mechanism}
        \end{figure}
        
        For model-annotated files, on the other hand, developers are presented with a visual summary of the distribution of model-annotated reviews as soon as the annotation task is complete  (\autoref{fig:developer_dash}). SENSOR also features an integrated feedback mechanism  (\autoref{fig:feedback_mechanism}) that enables developers to express disagreements with model-assigned labels for any particular review. Developers can access this mechanism by clicking on the ``\textit{eye}'' icon  (\autoref{fig:developer_dash}) corresponding to a model-annotated file on their dashboard. They can select the type of reviews (\textit{privacy-related} or \textit{privacy-irrelevant}) that they want to provide feedback on, from a drop-down list  (\autoref{fig:feedback_mechanism}). They can either download the file without any changes just by clicking on the ``\textit{Download}'' button or select the ``\textit{Load Reviews}'' button to provide their feedback. After the reviews have been loaded, they can check the corresponding boxes in the ``\textit{Disagree}'' column for annotations they disagree with (if any). On clicking the ``\textit{Submit Feedback and Download}'' button, their feedback will be recorded, and only the reviews without disagreement will be downloaded as a CSV file. This approach ensures that model assistance remains subject to human oversight. These structured mechanisms of monitoring progress and providing feedback encourage annotator awareness while enhancing transparency, accuracy, and reliability of the labeled datasets.

    \section*{Research Methodology}
        In this study, we obtained user reviews for seven popular social media applications on the Google Play Store based on their global impact, extensive feature sets, and the diversity of associated privacy challenges \cite{jain2021online, page2022social, miller2022don}. These applications included Instagram, Facebook, WhatsApp, Snapchat, X (formerly Twitter), Facebook Lite, and Line. By including these seven apps, we aim to capture not only the scale and diversity of user interactions across top global platforms but also to understand how distinct privacy features and regional contexts shape users' privacy concerns. This diversity enriches our analysis and improves the generalizability of our findings. This section describes our approach to scraping, filtering, and labeling user reviews using the SENSOR tool, followed by pre-processing, augmentation, and post-processing steps to generate training, testing, and validation datasets for training and evaluating various classifiers. We also elaborate on the proposed deep-learning architecture, \textbf{GRACE} (\textbf{GR}U-based \textbf{A}ttention with \textbf{C}BOW \textbf{E}mbedding), alongside the other models that were evaluated for classifying user reviews either as --- \textit{privacy-related feature requests} or \textit{privacy-related bug reports}, or \textit{privacy-irrelevant}.

    \subsection*{Scrapping, Filtering, and Labeling of User Reviews}    
        Due to the ubiquitous usage of social media applications, the volume and diversity of user-generated app reviews have grown significantly in recent times \cite{hatamian2019revealing,nema2022analyzing}. Although the literature suggests that only 0.5\% of these reviews are privacy-related \cite{mukherjee2020empirical}, filtering them out from a huge user-review base, followed by manual annotation, is a tedious and time-consuming task \cite{ebrahimi2022unsupervised,hatamian2019revealing}. To ease this situation, researchers have previously considered user reviews greater than or equal to a certain length \cite{ebrahimi2022unsupervised} and employed a keywords-based approach to filter out candidate user reviews that potentially contain privacy concerns, followed by manual annotation \cite{mukherjee2020empirical, nguyen2019short, ebrahimi2022unsupervised, hatamian2019revealing, tao2020identifying} for training various machine learning models to automate the process of annotation \cite{besmer2020investigating}.  
        
        In this study, we utilized the user-review-scraping feature of the SENSOR tool to scrape a total of 78000 reviews from the Google Play Store across the seven applications, including 18000 reviews from Facebook and 10000 reviews from each of the remaining applications. Similar to prior studies \cite{mukherjee2020empirical, nguyen2019short, ebrahimi2022unsupervised, tao2020identifying}, we considered user reviews with at least \textit{five} words and filtered out potential privacy-related user reviews using a predefined set of keywords, grouped under the five themes --- \textit{privacy and data security}, \textit{access control and permissions}, \textit{account and user management}, \textit{security threats and issues}, and \textit{tracking and monitoring}. Most of these keywords, as listed in \autoref{tab:privacy_keywords}, were adopted from the literature \cite{mukherjee2020empirical, nguyen2019short, ebrahimi2022unsupervised, tao2020identifying}. Although these keywords helped ease the process of filtering privacy-related reviews, we were mindful that some relevant reviews might not include any of those keywords. To mitigate this risk, we conducted a manual inspection of a random subset of keyword-excluded reviews, which revealed a small number of privacy-relevant reviews that used implicit language or indirect references. These insights informed an iterative refinement of the keywords list (\autoref{tab:privacy_keywords}), and several such terms were added to improve coverage.

        \begin{table}[htbp]
            \centering
            \scriptsize
            \caption{Privacy-related keywords grouped under five themes.}
            \begin{tabularx}{1\textwidth}{L{36ex}L{85ex}}
                \toprule
                \parbox{40ex}{\centering \textbf{Theme}} & \parbox{85ex}{\centering \textbf{Keywords}}\\
                
                \midrule
                Privacy and Data Security & \textit{privacy, security, personal, information, data, encrypt, encryption, anonymous, protect, protection, breach, leak, disclose, disclosure, unauthorized, identity, authentication, verification, consent, policy, terms, conditions.} \\\\
                
                Access Control and Permissions & \textit{access, control, permissions, restrict, restricted, authorized, unauthorized, approve, deny, allow, revoke, manage, settings, preferences, visibility, public, private, shared, share, blocked, ban.} \\\\
                
                Account and User Management & \textit{account, login, logout, log in, log out, recover, recovery, deletion, delete, remove, removed, deactivate, deactivate, disabled, enabled, locked, deactivate, request, review, appeal, notify, notification, history, activity.}\\\\
                
                Security Threats and Issues & \textit{hack, hacked, compromised, compromise, expose, exposed, failure, fail, failed, problem, issue, error, glitch, bug, crash, stuck, malfunction, unintended, unsafe, insecure, missing, disappeared, lost.}\\\\
                
                Tracking and Monitoring & \textit{tracking, track, location, visibility, notify, notification, audit, log, reviewed, retain, customize, report.}\\
        
                \bottomrule
            \end{tabularx}
            \label{tab:privacy_keywords}
        \end{table}

        Using these keywords, we filtered out approximately 12170 reviews, potentially related to privacy concerns (\textit{feature requests} or \textit{bug reports}). Additionally, we randomly selected 3830 unique \textit{privacy-irrelevant} reviews without imposing any limit on the review length from the full dataset of 78000 reviews, resulting in a final set of 16000 user reviews across the seven social media applications for manual annotation, the distribution of which is detailed in \autoref{tab:review_stat}.
        \begin{table}[htbp]
            \centering
            \scriptsize
            \caption{Statistics of reviews scraped and candidate reviews selected for annotation across seven social media applications.}
            \begin{tabularx}{.89\textwidth}{C{25ex}C{25ex}C{45ex}}
                \toprule
                \textbf{Application} & \textbf{Reviews Scraped} & \textbf{Candidate Reviews for Annotation}\\
                
                \midrule
                Instagram & 10000 & 1688 \\
                Facebook & 18000 & 5136 \\
                Whatsapp & 10000 & 1362 \\
                Snapchat & 10000 & 1588 \\
                X (formerly Twitter) & 10000 & 2007 \\
                Facebook Lite & 10000 & 643 \\
                Line & 10000 & 3576 \\
                \hline
                Total & 78000 & 16000\\
                \bottomrule
            \end{tabularx}
            \label{tab:review_stat}
        \end{table}
        
        The first two authors manually categorized each of the 16000 reviews into one of three categories --- \textit{privacy-related feature requests}, \textit{privacy-related bug reports}, or \textit{privacy-irrelevant}. They independently annotated the dataset using the manual annotation feature of the SENSOR tool, which resulted in a Cohen’s Kappa coefficient of 0.87, reflecting a high inter-rater agreement \cite{rau2021evaluation, wang2022coolted}. Any discrepancies in labeling were resolved through discussion between the two authors to reach a consensus. In the annotated dataset, there were 3627 \textit{privacy-related feature requests}, 4221 \textit{privacy-related bug reports}, and 8152 \textit{privacy-irrelevant} reviews.
              
    \subsection*{Data Pre-processing and Augmentation}
        Text pre-processing plays a vital role in minimizing noise and eliminating redundant information from user reviews, thereby improving the performance (precision and recall) of various classifiers \cite{qureshi2021performance, reddy2024classification}. Furthermore, data augmentation is an established technique for improving the generalizability of machine learning models by expanding the training dataset \cite{feng2021survey}. 
        
        In this study, the text pre-processing process involved several key steps, including --- \textit{converting text to lowercase}, \textit{expanding contractions} (e.g., ``don’t'' to ``do not'') \cite{tao2020identifying}, removing \textit{HTML elements} \cite{besmer2020investigating}, \textit{numbers}, \textit{usernames}, \textit{extra spaces}, \textit{special characters}, \textit{duplicate reviews}, and \textit{empty reviews}. After pre-processing the 16000 user reviews, the dataset was split into training, validation, and test sets using an 80-10-10 split, resulting in 12756, 1594, and 1595 samples, respectively. 
        
        The training set was then augmented using several NLP techniques using the \texttt{nlpaug} library from \textit{python}, including \textit{random word drop}, \textit{synonym substitution}, \textit{contextual word substitution}, \textit{contextual word insertion}, and \textit{abstract summarization} \cite{joshi2023text}. It is to be noted that punctuation marks such as \textit{commas}, \textit{semicolons}, \textit{periods}, \textit{question marks}, and \textit{exclamation points} were retained during pre-processing, allowing the augmentation models to capture the sentence-level context while ensuring a lexically diverse training set of enhanced quality. Each augmentation technique was applied to each of the reviews in the training set a specific number of times (as detailed in \autoref{tab:review_distribution}), resulting in a pre-processed augmented training dataset containing 126602 samples, including the original 12756 samples.

        \begin{table}[htbp]
            \centering
            \scriptsize
            \caption{Distribution of reviews in the augmented training set across different classes (\textit{privacy-related feature request}, \textit{privacy-related bug report}, and \textit{privacy-irrelevant}), including those in the pre-processed training set and those generated using various data augmentation techniques.}
            \begin{tabularx}{1\textwidth}{C{35ex}C{20ex}C{13ex}C{13ex}C{13ex}C{10ex}}
                \toprule
                \multirow{2}{*}{\parbox{35ex}{\centering \textbf{Dataset / \\Augmentation Technique}}} & 
                \multirow{2}{*}{\parbox{20ex}{\centering \textbf{Augmentation per Review}}} & 
                \multicolumn{3}{c}{\parbox{40ex}{\centering \textbf{Augmented Reviews per Class} }} & 
                \multirow{2}{*}{\parbox{10ex}{\centering \textbf{Total}}}\\
                \cline{3-5}
                & & \parbox{13ex}{\centering \textbf{PFR$^a$}} 
                & \parbox{13ex}{\centering \textbf{PB$^b$}}
                & \parbox{13ex}{\centering \textbf{PIR$^c$}} &\\
                \midrule
                Pre-processed Training Set & - & 2878 & 3366 & 6512 & 12756\\
                Random Word Drop & 2 & 5696 & 6653 & 12674 & 25023\\
                Synonym Substitution & 2 & 5719 & 6669 & 12792 & 25180\\
                Contextual Word Substitution & 2 & 5755 & 6728 & 12987 & 25470\\
                Contextual Word Insertion & 2 & 5748 & 6727 & 12957 & 25432\\
                Abstract Summarization & 1 & 2874 & 3362 & 6505 & 12741\\
                \hline
                \multicolumn{2}{r}{\textbf{Total Pre-processed Augmented Reviews}} & 28670 & 33505 & 64427 & 126602\\
                \multicolumn{2}{r}{\textbf{Total Post-processed Augmented Reviews}} & \textbf{27485 (22.64\%)} & \textbf{32320 (26.63\%)} & \textbf{61569 (50.73\%)} & \textbf{121374}$^*$\\
                
                \bottomrule
                \multicolumn{6}{l}{$^a$ Privacy-related Feature Request (PFR)}\\
                \multicolumn{6}{l}{$^b$ Privacy-related Bug (PB)}\\
                \multicolumn{6}{l}{$^c$ Privacy-irrelevant (PIR)}\\
                \multicolumn{6}{l}{$^*$ Final Training Set}
            \end{tabularx}
            \label{tab:review_distribution}
        \end{table}
        
        \begin{figure}[t]
            \centering
            [FIGURE \ref{fig:data_pre-processing}]
            \vspace{5mm}
            \includegraphics[width=\textwidth]{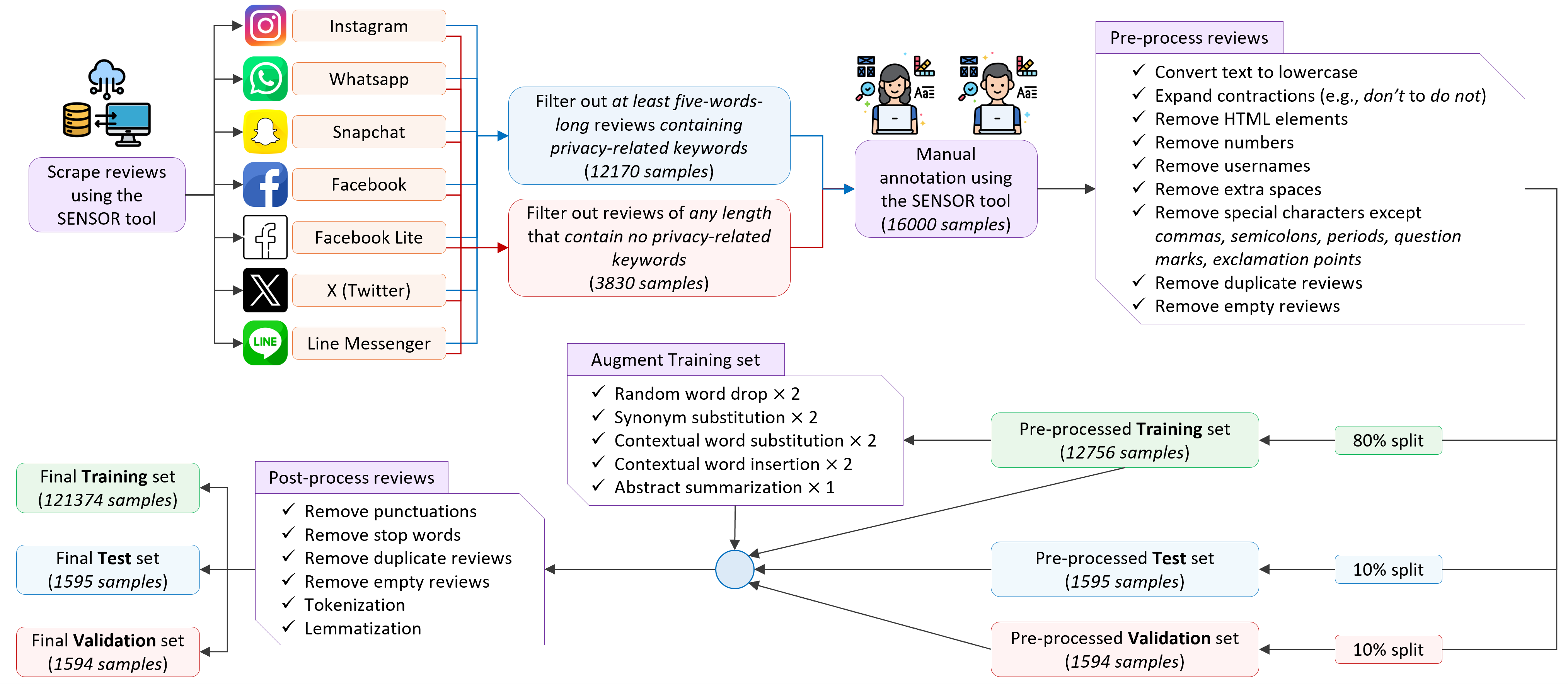}
            \caption{\textbf{Workflow diagram showing the steps involved in generating the processed train, validation, and test sets of user reviews, including --- scraping, filtering, annotation, pre-processing, augmentation, and post-processing, for training various classifiers.}}
            \label{fig:data_pre-processing}
        \end{figure}
        
        The augmented training set, along with validation and test sets, underwent several post-processing steps, including the removal of --- \textit{duplicate reviews}, \textit{empty reviews}, \textit{punctuation marks}, and \textit{stop words} (based on the list provided in the \texttt{nltk} library \cite{bird2006nltk}) to reduce noise and unwanted information \cite{qureshi2021performance, Maalej2016OnTA}. Finally, \textit{tokenization} and \textit{lemmatization} were applied to refine the dataset further \cite{qureshi2021performance, hatamian2019revealing, besmer2020investigating, Maalej2016OnTA}, resulting in an augmented final training dataset of 121374 samples (\autoref{tab:review_distribution}), which included 27485 \textit{privacy-related feature requests}, 32320 \textit{privacy-related bug reports}, and 61569 \textit{privacy-irrelevant} reviews for training and evaluation of various machine learning models. \autoref{fig:data_pre-processing} illustrates the workflow of generating the train, validation, and test sets for training machine learning models, including scraping, filtering, annotating, pre-processing, augmenting, and post-processing of the user reviews.

    \subsection*{User Review Classification}
        This section delves into the architecture of the proposed automated annotation model, GRACE, followed by an overview of other contemporary models considered in this study, including traditional and state-of-the-art deep learning techniques. We also discuss our approach to evaluating models' performance in classifying user reviews.
        
                    
        \subsubsection*{Developing and Training the Annotation Model: GRACE}
            In this paper, we developed an automated annotation model, \textbf{GRACE} (\textbf{GR}U-based \textbf{A}ttention with \textbf{C}BOW \textbf{E}mbedding), leveraging Gated Recurrent Units (GRU) \cite{dey2017gate} integrated with Continuous Bag of Words (CBOW) \cite{xia2023continuous} and the Attention Mechanism \cite{brauwers2021general} to classify user reviews into one of the three categories --- \textit{privacy-related feature request}, \textit{privacy-related bug}, and \textit{privacy-irrelevant}. The model, designed to handle 150-word-long textual input sequences, first transforms each token in the sequence into a dense 200-dimensional vector using an embedding layer that uses pre-trained weights from a CBOW-based embedding matrix trained on the processed training set. This allows the architecture to learn meaningful word representations. These embeddings are then passed into a GRU layer with 896 hidden units, which excels at capturing temporal dependencies in the sequence while keeping the computation relatively efficient. To enhance the model's ability to focus on the most relevant parts of the sequence, an attention mechanism is applied, the outputs of which are combined with those of the GRU layer to preserve the overall context and the attention-driven refinements. This combined representation is then summarized into a single vector by global average pooling (regularized with 50\% dropout to prevent overfitting) and refined using a dense layer with 256 units. Finally, an output layer applies \textit{softmax} activation on the outputs of this dense layer to produce a probability distribution over the target classes --- \textit{privacy-related feature request}, \textit{privacy-related bug}, and \textit{privacy-irrelevant}.

            The model was compiled using the Adam optimizer \cite{rahman2022two}, \textit{categorical cross-entropy} as the loss function, and \textit{accuracy} as the evaluation metric. It was then trained for 50 epochs with a batch size of 256 and an early-stopping mechanism with a patience value of three. In the event of early stopping, the best-performing weights were restored to prevent overfitting and improve generalization, while monitoring validation loss. 
            
        \subsubsection*{Training Classical and State-of-the-art Deep-Learning Framewoks}
            In addition to our proposed framework, GRACE, we conducted a comprehensive evaluation of both foundational machine learning algorithms and contemporary deep learning architectures by categorizing our methods into three distinct groups. The first group encompassed classical models, including Support Vector Machines, Logistic Regression, XGBoost, and LightGBM. These models were chosen for their established strengths in reliability, computational efficiency, and interpretability. Serving as robust baselines, they provided valuable benchmarks for assessing model performance on the text classification task and offered clear insight into their respective strengths and limitations.
            
            The second group focused on Recurrent Neural Network (RNN) architectures such as Long Short-Term Memory (LSTM), Gated Recurrent Unit (GRU), and Bidirectional LSTM (BiLSTM). To enhance their ability to model sequential dependencies and contextual semantics, we experimented with a variety of hybrid configurations, integrating these architectures with Continuous Bag-of-Words (CBOW) embeddings \cite{xia2023continuous} and the Attention Mechanism \cite{brauwers2021general}. Notable combinations included BiLSTM + CBOW, BiLSTM + CBOW + Attention, GRU + Attention, and GRU + CBOW, each designed to capture complex linguistic patterns across textual sequences more effectively.
            
            The third group featured transformer-based language models, namely BERT, DistilBERT, and RoBERTa,  which leverage large-scale pretraining to recognize and model deep contextual relationships within the text. These models represent the state-of-the-art in natural language processing and provide a powerful comparative benchmark for our framework.
            
            All deep learning experiments, including both RNN-based and transformer-based models, were implemented using TensorFlow or PyTorch. We employed \textit{categorical cross-entropy} as the loss function and optimized model performance through techniques such as early stopping, dropout regularization, and the Adam optimizer \cite{rahman2022two}, ensuring both accuracy and generalizability across evaluation scenarios.

        \subsubsection*{Model Evaluation}
            The evaluation process focuses on both individual class performance and overall classification accuracy, providing a detailed view of how each model handles the classification task. The performance of the proposed GRACE framework has been juxtaposed with those of all the machine learning models mentioned earlier, utilizing various evaluation metrics. The metrics included Precision, Recall, F1-Score, ROC-AUC, and Accuracy, which provided a comprehensive understanding of each model's classification accuracy. Metric-wise macro-averaged values were also analyzed to give insights into the generalization capability of a model across all classes. 

            Since the best-performing model will be deployed in an online environment for automated annotation tasks, it is also essential to evaluate its operational efficiency. The size of the model and the inference time are critical factors in ensuring a smooth, scalable deployment. A smaller model size reduces storage requirements and minimizes resource consumption, which is vital for maintaining efficient operations, especially in resource-constrained environments. Inference time, on the other hand, directly impacts the system's responsiveness and the ability to process annotations in real time, ensuring swift annotation without delays. By incorporating these practical aspects alongside traditional performance metrics, we get an overall insight into each model’s suitability for production use. This holistic approach enables us to select a model that excels both in classification and meets the operational demands of online deployment, balancing accuracy with responsiveness and resource efficiency to deliver both high-quality results and a seamless user experience.

\section*{Results}    

    \subsection*{Classifier Performance Analysis}
        \begin{figure}[!b]
            \centering
            [FIGURE \ref{fig:classswise_HSGD}] 
            \vspace{5mm}
            
            \subfloat[]{
                \includegraphics[width=.482\textwidth]{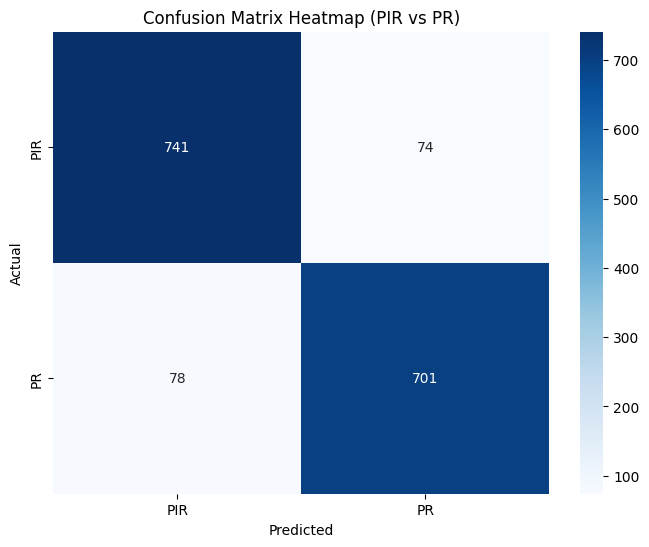}%
                \label{fig:conf_mat_PIR_PR_HSGD}
            }
            \hfill
            \subfloat[]{
                \includegraphics[width=.482\textwidth]{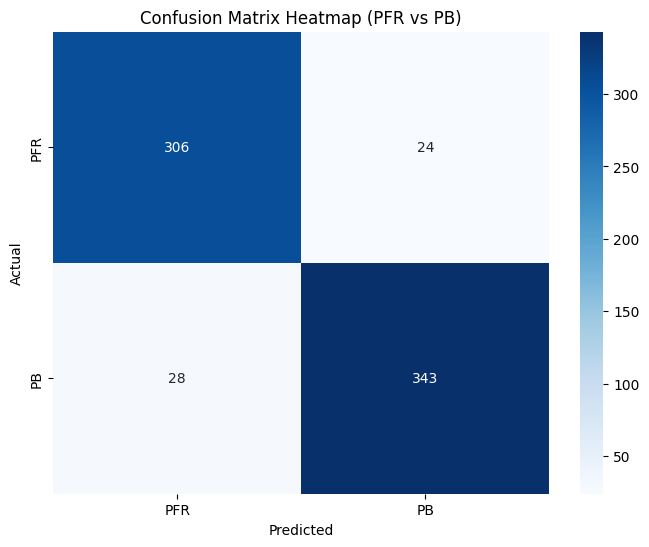}%
                \label{fig:conf_mat_PFR_PB_HSGD}
            }
            
            \subfloat[]{
                \includegraphics[width=.482\textwidth]{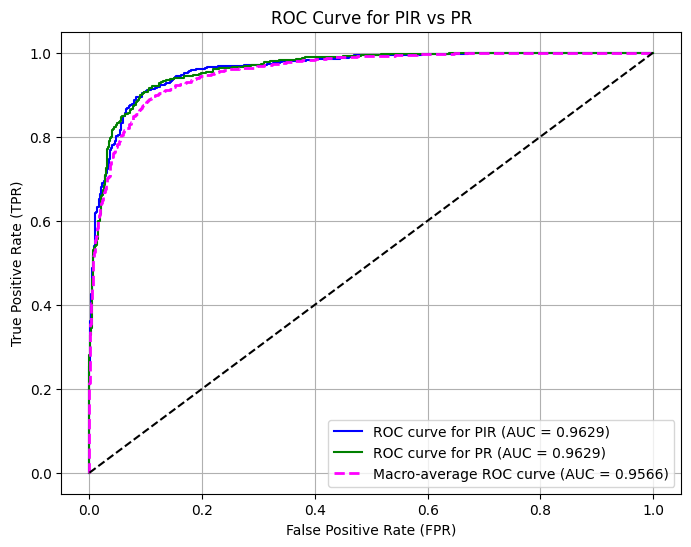}%
                \label{fig:roc_auc_PIR_PR_HSGD}
            }
            \hfill
            \subfloat[]{
                \includegraphics[width=.482\textwidth]{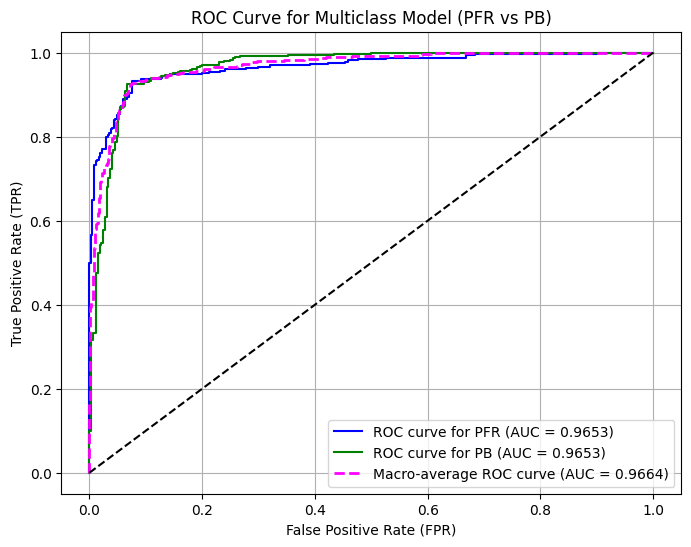}%
                \label{fig:roc_auc_PFR_PB_HSGD}
            }
            \caption{\textbf{Confusion matrices (a–b) and ROC-AUC plots (c–d) for the Hierarchical SGD model in classifying user reviews as: (a, c) \textit{privacy-related} (PR) vs. \textit{privacy-irrelevant} (PIR), and (b, d) \textit{privacy-related feature request} (PFR) vs. \textit{privacy-related bug report} (PB).
}}
            \label{fig:classswise_HSGD}
        \end{figure}
        
        As benchmarks, classical machine-learning models, including Logistic Regression (SGD with \textit{log-loss} loss function), linear SVM (SGD with \textit{hinge} loss), XGBoost, and LightGBM, yielded modest results using combinations of TF-IDF \cite{qureshi2021performance} and word2vec \cite{church2017word2vec} as text representation methods. The macro-F1 scores ranged from 0.7800 to 0.8500, while the macro-ROC-AUC values ranged from 0.9261 to 0.9576. For all these classical models, the F1-score and the ROC-AUC for the class \textit{privacy-irrelevant} were consistently the highest. This suggests that classical machine-learning classifiers are most effective at recognizing \textit{privacy-irrelevant} content, likely because such reviews have more distinctive, non-privacy vocabulary, and/or form the majority of the data while struggling more to distinguish between the more nuanced \textit{privacy-related} classes. This observation led us to evaluate a hierarchical SGD variant where the reviews were first classified as either \textit{privacy-related} or \textit{privacy-irrelevant}. The filtered-out \textit{privacy-related} reviews were then further classified as \textit{feature requests} or \textit{bugs}. This hierarchical structure outperformed the other classical models, achieving a macro-F1 score of 0.86 with an accuracy of 87.20\%. The macro ROC-AUC of differentiating between \textit{privacy-related} and \textit{privacy-irrelevant} was 0.9566 (\autoref{fig:roc_auc_PIR_PR_HSGD}), with the confusion matrix shown in \autoref{fig:conf_mat_PIR_PR_HSGD}, and that between \textit{privacy-related feature request} and \textit{privacy-related bug} was 0.9664 (\autoref{fig:roc_auc_PFR_PB_HSGD}), with the confusion matrix shown in \autoref{fig:conf_mat_PFR_PB_HSGD}, suggesting the benefits of structured optimization in this domain. However, all classical methods were constrained by their fixed feature spaces and inability to model long-range dependencies, requiring robust feature representations.

        \begin{figure}[ht]
            \centering
            [FIGURE \ref{fig:classwise_GRACE}] 
            \vspace{5mm}
            
            \subfloat[]{
                \includegraphics[width=.48\textwidth]{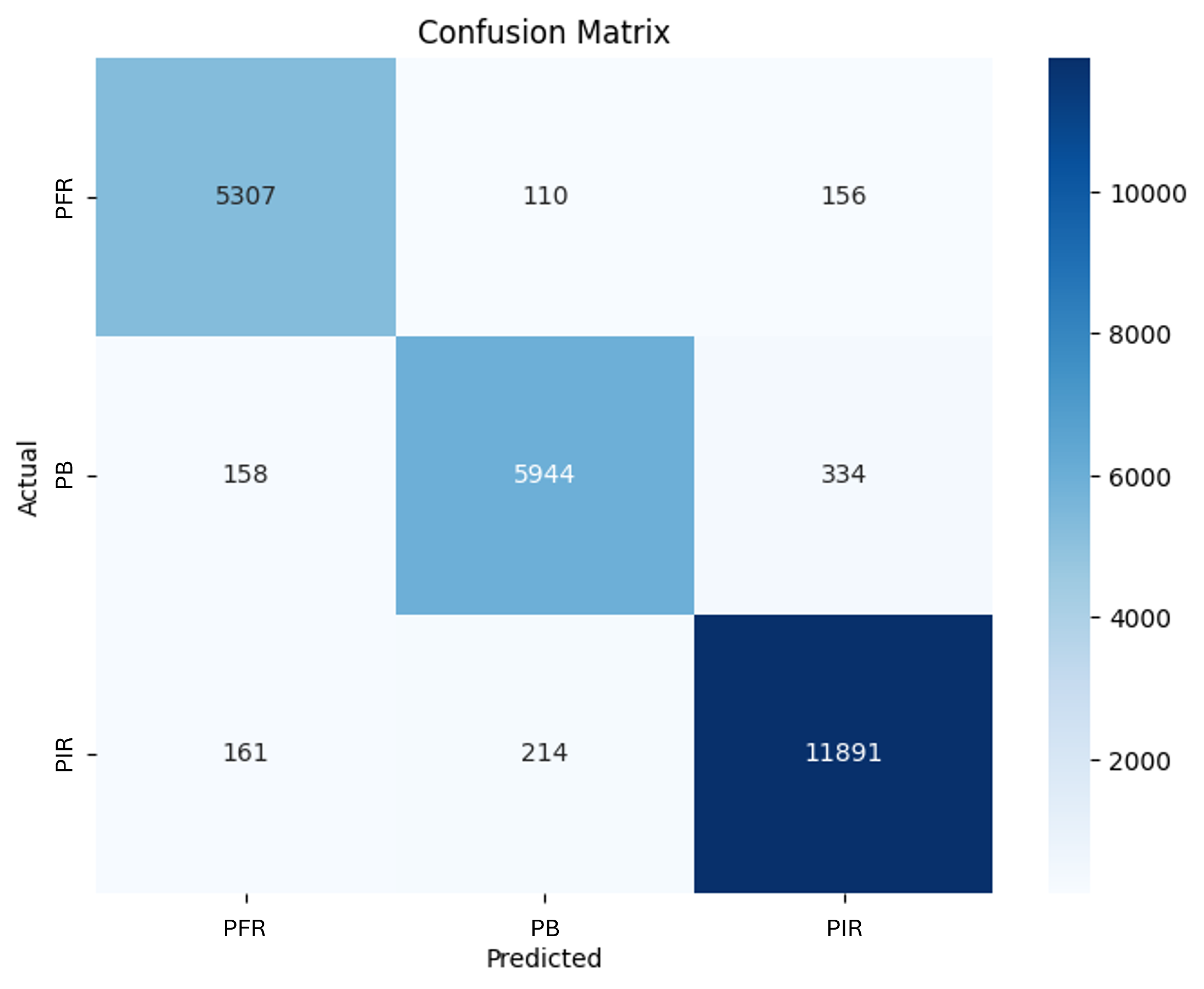}%
                \label{fig:conf_mat_grace}
            }
            \hfill
            \subfloat[]{
                \includegraphics[width=.48\textwidth]{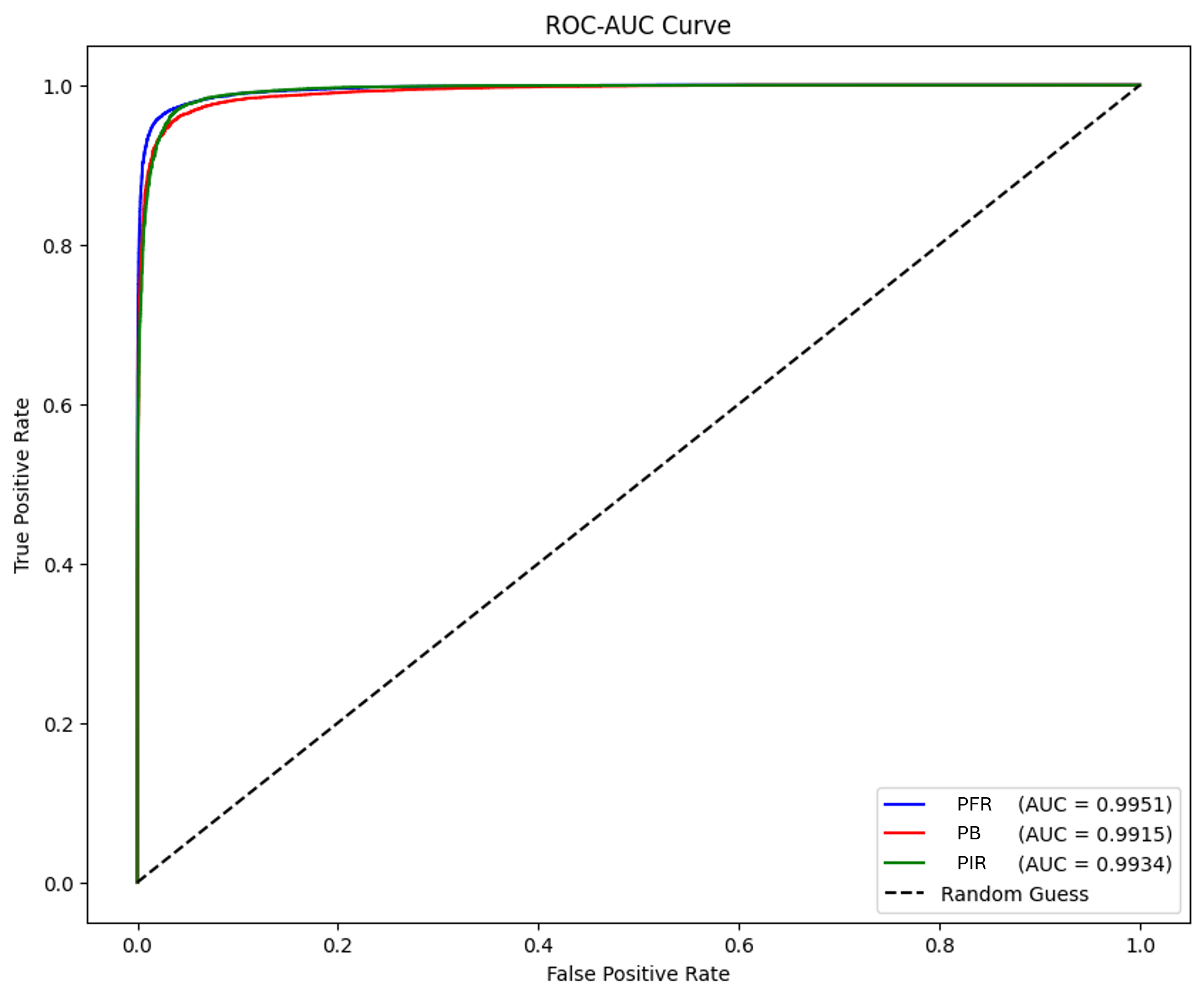}%
                \label{fig:roc_auc_grace}
            }
            \caption{\textbf{(a) Confusion matrix heatmap and (b) ROC-AUC plot of the GRACE framework in classifying user reviews as \textit{privacy-related feature request} (PFR), \textit{privacy-related bug reports} (PB), or \textit{privacy-irrelevant} (PIR).}}
            \label{fig:classwise_GRACE}
        \end{figure}
        
        Among the transformer-based models, RoBERTa, due to its heavier pretraining, struck a compelling balance of capacity and efficiency (macro-F1 score: 0.8712, macro ROC-AUC: 0.9671, accuracy: 88.08\%), while BiLSTM rendered a similar performance with slightly higher accuracy than RoBERTa (macro-F1 score: 0.8674, macro ROC-AUC: 0.9654, accuracy: 88.14\%), underscoring the joint benefits of bidirectional context and dynamic token weighting. 

        \begin{table}[htbp]
            \centering
            \scriptsize
            \caption{Macro-averaged values of evaluation parameters for various machine-learning models evaluated in this study.}
            \setlength{\tabcolsep}{6pt}
            \begin{tabularx}{1\textwidth}{C{40ex}C{12ex}C{10ex}C{12ex}C{15ex}C{13ex}}
                \toprule
                \textbf{Architecture} & \textbf{Precision} & \textbf{Recall} & \textbf{F1-Score} & \textbf{ROC-AUC} & \textbf{Accuracy (\%)}\\
                \midrule

                \textbf{GRACE$^a$} & \textbf{0.9483} & \textbf{0.9389} & \textbf{0.9434} & \textbf{0.9934} & \textbf{95.10}\\
                BiLSTM & 0.8743 & 0.8620 & 0.8674 & 0.9654 & 88.14\\
                RoBERTa & 0.8688 & 0.8745 & 0.8712 & 0.9671 & 88.08\\
                BiLSTM + CBOW$^b$ + Attention & 0.8636 & 0.8632 & 0.8634 & 0.9667 & 87.45\\
                Hierarchichal SGD & 0.8600 & 0.8600 & 0.8600 & 0.9416 & 87.20\\
                DistilBERT & 0.8569 & 0.8656 & 0.8609 & 0.9673 & 87.14\\
                GRU + Attention & 0.8532 & 0.8467 & 0.8474 & 0.9643 & 86.70\\
                BiLSTM + CBOW & 0.8506 & 0.8607 & 0.8550 & 0.9610 & 86.64\\
                Linear SVM & 0.8500 & 0.8500 & 0.8500 & 0.9576 & 86.39\\
                Logistic Regression & 0.8500 & 0.8500 & 0.8500 & 0.9568 & 86.20\\
                BERT & 0.8349 & 0.8542 & 0.8432 & 0.9586 & 85.38\\
                GRU & 0.8393 & 0.8426 & 0.8399 & 0.9559 & 85.38\\
                GRU + CBOW & 0.8380 & 0.8227 & 0.8275 & 0.9488 & 84.50\\
                LSTM & 0.8266 & 0.8382 & 0.8317 & 0.9426 & 84.32\\
                XGBoost & 0.8100 & 0.7900 & 0.8000 & 0.9481 & 82.37\\
                LightGBM & 0.7800 & 0.7900 & 0.7800 & 0.9261 & 79.99\\
                RNN & 0.7596 & 0.7724 & 0.7650 & 0.9185 & 78.86\\
                
                \bottomrule\\
                \multicolumn{6}{l}{$^a$ Proposed architecture.}\\
                \multicolumn{6}{l}{$^b$ Continuous Bag of Words (CBOW).}\\
                
            \end{tabularx}
            \label{tab:model_evaluation}
        \end{table}
        
        The proposed \textbf{GRACE} framework, which integrates Gated Recurrent Units (GRU) \cite{dey2017gate}, Continuous Bag of Words (CBOW) \cite{xia2023continuous}, and the Attention Mechanism \cite{brauwers2021general}, achieved the best overall performance. It achieved a macro-average precision of 0.9483, recall of 0.9389, F1-score of 0.9434, and ROC-AUC of 0.9934, corresponding to 95.10\% overall accuracy. The confusion matrix and the ROC-AUC plot of the GRACE framework are illustrated in \autoref{fig:conf_mat_grace} and \autoref{fig:roc_auc_grace}, respectively. These gains can be attributed to the CBOW layer’s robust representations for infrequent tokens, the GRU’s gating mechanisms that mitigate vanishing gradients, and the attention module’s capacity to highlight salient sequence elements dynamically. Moreover, a moderate dropout rate (0.5) and early stopping (patience = 3) helped prevent overfitting despite the model’s high parameter count. \autoref{tab:model_evaluation} summarizes the performance evaluation of all the models in combination with various text representation techniques that were considered in this study.
        
\subsection*{Inference Time Analysis}
        The comparison of models reveals notable trade-offs between inference efficiency and model size. The proposed GRACE framework achieved the highest accuracy (95.10\%) with a moderate model size of 44.09 MB and a mean inference time of 115.03 ms, highlighting a strong balance between performance and computational demands. In contrast, transformer-based models like RoBERTa and BERT, while maintaining competitive accuracies (88.08\% and 85.38\% respectively), exhibited significantly larger model sizes (475.58 MB and 417.73 MB) and varied inference times. RoBERTa, for example, had a notably low mean inference time of 26.28 ms, whereas BERT required 28.39 ms despite its larger size. 
        
        \begin{table}[htbp]
            \centering
            \scriptsize
            \caption{Statistics of model size and inference times compared to respective accuracy, sorted in descending order by accuracy.}
            \setlength{\tabcolsep}{6pt}
            \begin{tabularx}{1\textwidth}{C{40ex}L{12ex}L{10ex}L{12ex}L{12ex}L{13ex}}
                \toprule
                \multirow{2}{*}{\parbox{40ex}{\centering \textbf{Architecture}}} & 
                \multirow{2}{*}{\parbox{15ex}{\centering \textbf{Model Size (MB)}}} &
                \multirow{2}{*}{\parbox{10ex}{\centering \textbf{Accuracy (\%)}}} & 
                \multicolumn{3}{c}{\parbox{30ex}{\centering \textbf{Inference Times (ms)} }} \\
                \cline{4-6}
                & & & \parbox{15ex}{\centering \textbf{Min}} 
                & \parbox{15ex}{\centering \textbf{Max}}
                & \parbox{15ex}{\centering \textbf{Mean}}\\
                \midrule

                \textbf{GRACE$^a$} & \textbf{44.09} & \textbf{95.10} & \textbf{65.26} & \textbf{225.02} & \textbf{115.03}\\
                BiLSTM & 13.65 & 88.14 & 50.47 & 213.74 & 73.44 \\
                RoBERTa & 475.58 & 88.08 & 16.51 & 86.52 & 26.28\\
                BiLSTM + CBOW$^b$ + Attention & 52.64 & 87.45 & 47.58 & 97.17 & 68.83\\
                Hierarchichal SGD & 225.99 & 87.20 & 219.20 & 252.76 & 231.12\\
                DistilBERT & 255.46 & 87.14 & 9.99 & 85.00 & 14.90\\
                GRU + Attention & 40.87 & 86.70 & 61.21 & 364.54 & 77.96\\
                BiLSTM + CBOW & 26.05 & 86.64 & 43.38 & 495.22 & 59.23\\
                Linear SVM & 225.98 & 86.39 & 206.37 & 330.21 & 230.71\\
                Logistic Regression & 225.98 & 86.20 & 209.11 & 302.61 & 229.59\\
                BERT & 417.73 & 85.38 & 18.00 & 97.83 & 28.39\\
                GRU & 40.86 & 85.38 & 48.97 & 226.67 & 61.49\\
                GRU + CBOW & 3.00 & 84.50 & 51.45 & 635.86 & 69.92\\
                LSTM & 40.90 & 84.32 & 45.78 & 372.91 & 55.34\\
                XGBoost & 309.06 & 82.37 & 2076.68 & 18914.72 & 3597.31\\
                LightGBM & 291.76 & 79.99 & 217.63 & 422.03 & 249.67\\
                RNN & 18.73 & 78.86 & 58.36 & 1326.69 & 109.26\\
                
                \bottomrule\\
                \multicolumn{6}{l}{$^a$ Proposed architecture.}\\
                \multicolumn{6}{l}{$^b$ Continuous Bag of Words (CBOW).}\\
                
            \end{tabularx}
            \label{tab:size_inference_stat}
        \end{table}
        
        Classical models such as Logistic Regression and SVM showed lower accuracies (86.2–86.4\%) with relatively large model sizes (225.98 MB), without substantial gains in inference speed. Notably, XGBoost, while delivering modest accuracy (82.37\%), incurred the highest mean and maximum inference times (3597.31 ms and 18914.72 ms, respectively), rendering it computationally expensive. The vanilla RNN with the least model complexity (18.73 MB) and the lowest accuracy (78.86\%), had a relatively high maximum inference time (1326.69 ms), underscoring its limitations in both predictive power and efficiency. However, reducing model size alone did not guarantee speed; the GRU + CBOW (3 MB) variant processed inputs in approximately 70 ms on average, almost five times slower than the much larger DistilBERT (255.46 MB). 
        
        \begin{figure}[htbp]
            \centering
            [FIGURE \ref{fig:model_inference_size}]
            \includegraphics[width=\textwidth]{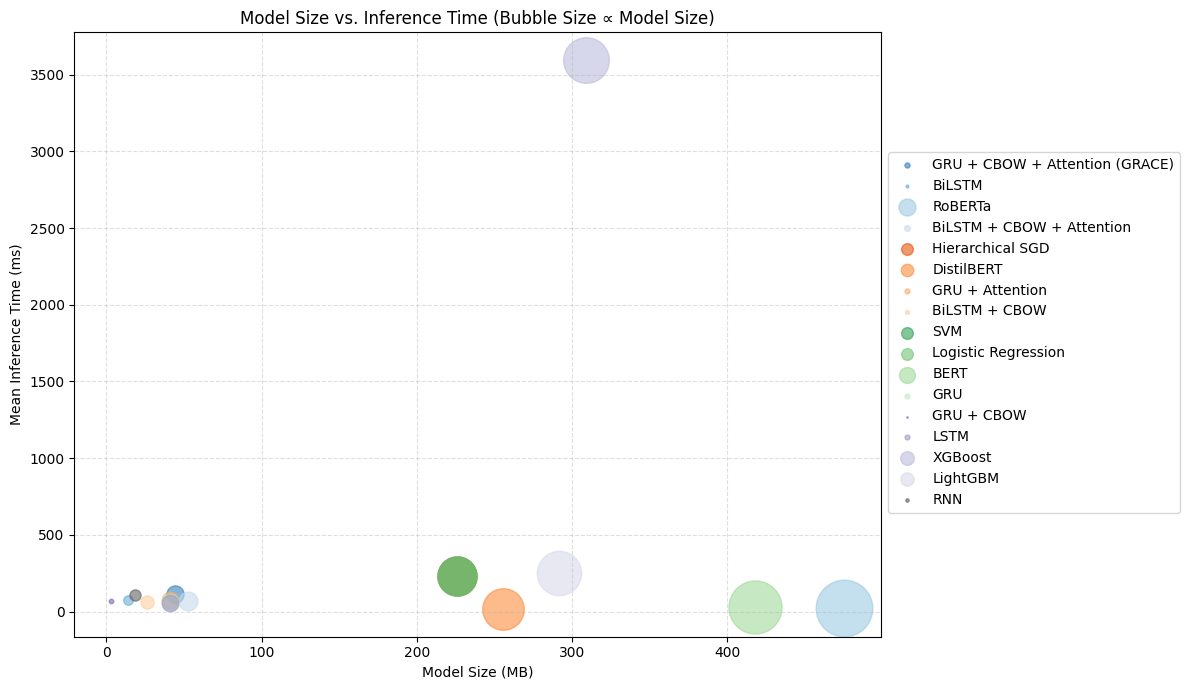}
            \caption{\textbf{Normalized comparison of model size and inference latency across architectures. Each shaded circle’s area scales with the original model size.}}
            \label{fig:model_inference_size}
        \end{figure}
        
        Overall, GRU-based architectures, particularly when enhanced with attention and embedding strategies, provide a compelling balance of accuracy, model compactness, and inference time, whereas distilled transformers are preferable when ultra-low latency outweighs memory considerations. \autoref{tab:size_inference_stat} statistically summarizes the inference times and model size compared to the respective accuracy. The bubble plot in \autoref{fig:model_inference_size} illustrates the comparison of model size and inference latency across the aforementioned models, where the area of each shaded circle scales with the original model size. 

        These results indicate that the GRACE framework occupies a favorable middle ground. It consistently outperforms classical and recurrent baselines, approaches the accuracy of distilled transformers, and retains a comparatively modest computational footprint. For practitioners who require strong performance but cannot afford the full cost of large-scale transformers, GRACE offers a compelling, accessible alternative.

\section*{Threats to Validity}
    To provide a transparent assessment of our methodology and findings, we outline key threats to the validity of this study across four dimensions in this section. First, we discuss the \textit{Loss of Lexical Diversity Post-Augmentation}, highlighting how augmentation may have reduced vocabulary richness and impacted model generalization. Next, we address concerns related to \textit{Internal and External Validity}, including annotation biases, keyword-based filtering limitations, and the constrained scope of our data sources. We then examine \textit{Construct Validity}, reflecting on the alignment between our predefined classes and the nuanced, often overlapping nature of user privacy concerns. Finally, we consider \textit{Conclusion Validity}, acknowledging the absence of statistical significance testing and real-world user feedback in evaluating the GRACE framework. Recognizing these limitations is essential to accurately interpreting our results and guiding future improvements.
    
    \subsection*{Loss of Lexical Diversity Post-Augmentation}
        An important contribution of this study is the development of a large, manually labeled dataset containing 16000 (3627 \textit{privacy-related feature requests}, 4221 \textit{privacy-related bug reports}, and 8152 \textit{privacy-irrelevant}) reviews with a high inter-rater agreement (Cohen’s Kappa = 0.87). This dataset not only served as a robust training ground for GRACE but also represents a valuable resource for future research into privacy-aware NLP. 
        
        Eighty percent of the reviews (12756 samples) from this dataset, considered as the training set, were later augmented and processed to generate the final augmented training set containing 121374 reviews (\autoref{fig:data_pre-processing}), with 27485 \textit{privacy-related feature requests}, 32320 \textit{privacy-related bug reports}, and 61569 \textit{privacy-irrelevant} reviews (\autoref{tab:review_distribution}). This dataset retained a naturally skewed distribution across the three classes. Therefore, it is worth noting that the data augmentation techniques applied in this study were not intended to address class imbalance in the dataset, but rather to enhance the model’s ability to generalize under imbalanced conditions in the real world. While this approach yielded notable performance gains, especially for minority classes, several threats to validity must be acknowledged. 
    
        \begin{figure}[ht]
            \centering
            [FIGURE \ref{fig:mtld_kde_violin}] 
            \vspace{5mm}
            
            \subfloat[]{
                \includegraphics[width=.482\textwidth]{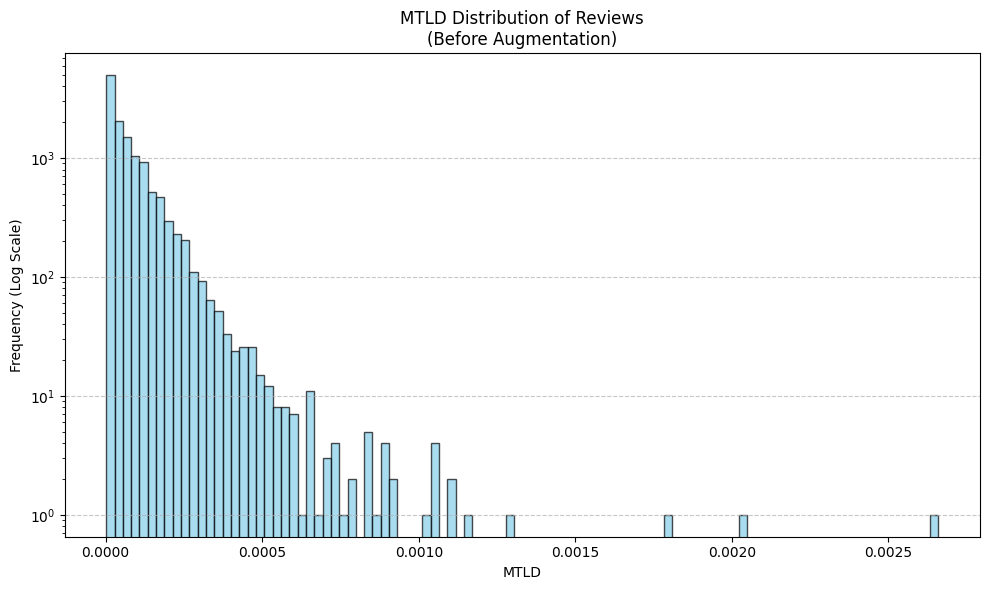}%
                \label{fig:mtld_raw}
            }
            \hfill
            \subfloat[]{
                \includegraphics[width=.482\textwidth]{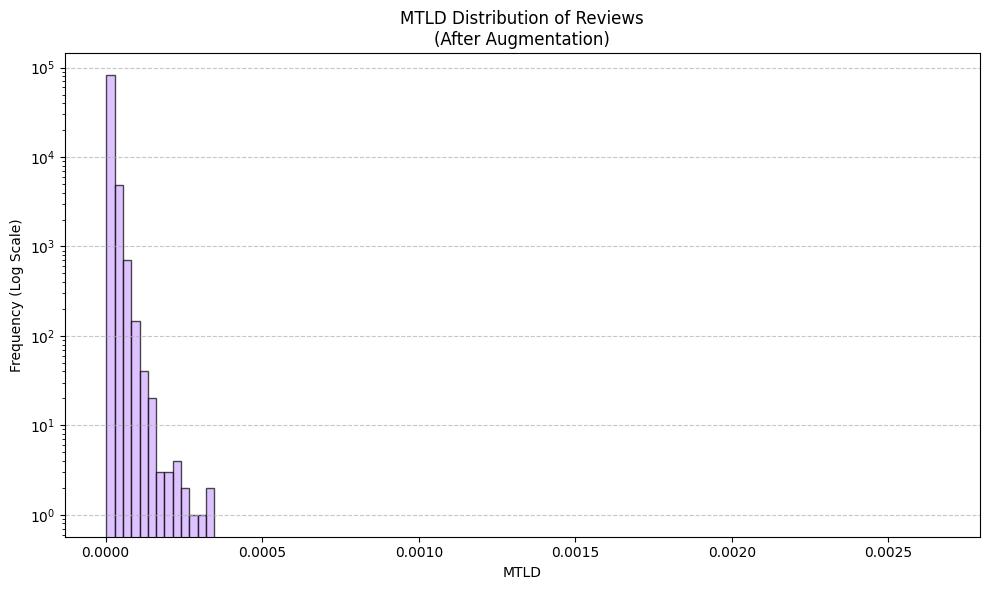}%
                \label{fig:mtld_postprocessed}
            }
            
            \subfloat[]{
                \includegraphics[width=.482\textwidth]{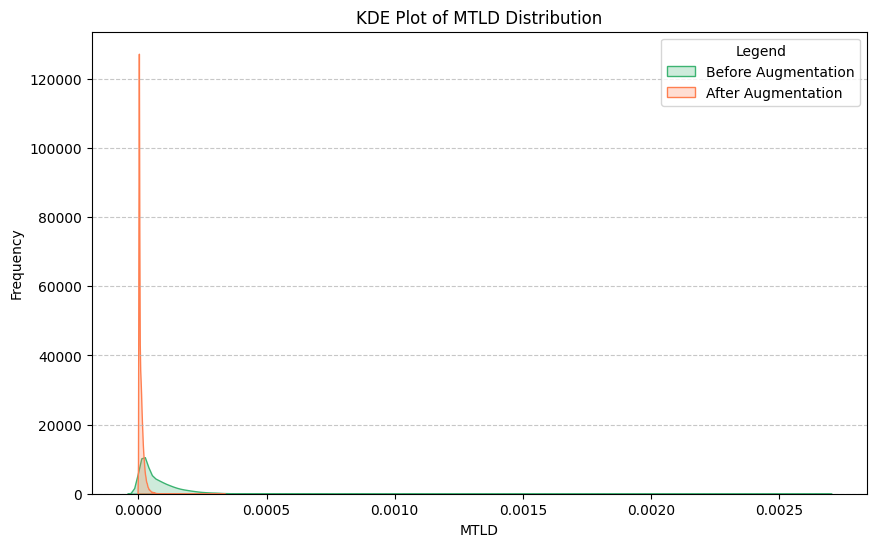}%
                \label{fig:kde}
            }
            \hfill
            \subfloat[]{
                \includegraphics[width=.482\textwidth]{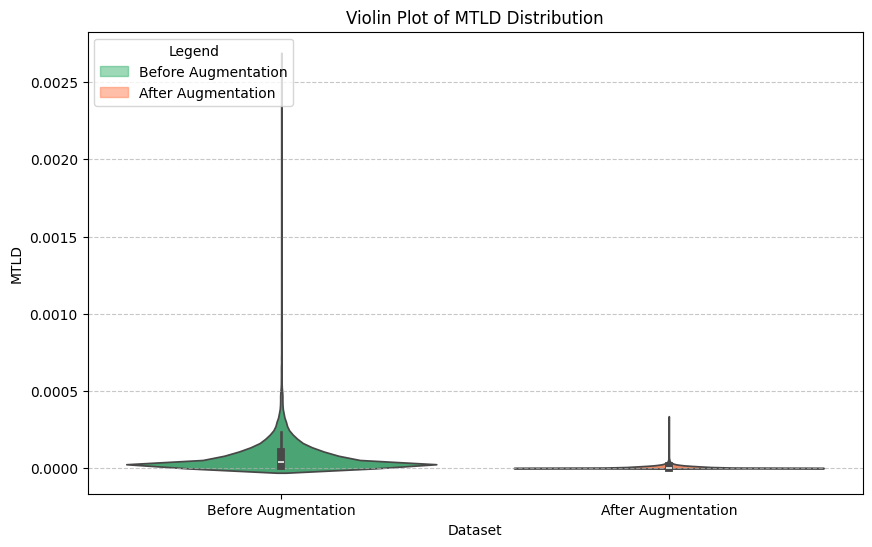}%
                \label{fig:violin}
            }
            \caption{\textbf{Distribution plot of the Measure of Textual Lexical Diversity (MTLD) of training set (a) before and (b) after augmentation. (c) KDE and (d) Violin plots of MTLD distribution of the training set before and after augmentation.}}
            \label{fig:mtld_kde_violin}
        \end{figure}
        
        A primary concern relates to the lexical diversity of the augmented dataset. If we compare the Kernel Density Estimate (KDE) \cite{sethuraman2021improved} plot (\autoref{fig:kde}) of the Measure of Textual Lexical Diversity (MTLD) (a metric used to assess the variance of the vocabulary within textual information) \cite{jarvis2021operationalizations} of the training set, before (\autoref{fig:mtld_raw}) and after (\autoref{fig:mtld_postprocessed}) augmentation, it can be seen that the lexical diversity of the training set exhibits a sharper and narrower peak after augmentation, compared to that before. This suggests that the augmented reviews are more lexically uniform, possibly due to repetitive phrasing and limited synonym sets employed in techniques such as synonym replacement and contextual word insertion. In contrast, the raw reviews demonstrate a broader distribution, suggesting greater variation in user expression. This reduction in diversity is further corroborated by the violin plot (\autoref{fig:violin}), where the augmented reviews are densely clustered around lower MTLD values. While such uniformity may help stabilize model training, the model may learn syntactic shortcuts or stylistic patterns rather than deeper semantic cues, which is a major concern in privacy-sensitive classifications, where subtle linguistic differences often differentiate \textit{privacy-related feature requests} from \textit{privacy-related bug reports} or \textit{privacy-irrelevant} content. Despite these drawbacks, the F1-score and ROC-AUC indicate that the augmentation strategy helped the GRACE framework remain robust in imbalanced settings. However, to further safeguard generalization and preserve lexical diversity, future work could explore semantic-aware or MTLD-preserving augmentation strategies that maintain diversity while avoiding label drift.

    \subsection*{Internal and External Validity}
        Our dataset was constructed using keyword-based filtering to identify privacy-related user reviews. Although we refined the keyword list through multiple iterations and manually inspected a subset of the excluded reviews, there remains a possibility that some genuinely privacy-relevant reviews were left out due to the absence of explicit keyword matches. These false negatives could introduce bias into the labeled dataset and may impact the model’s ability to generalize to broader, real-world inputs where privacy concerns are expressed more implicitly.
        
        Furthermore, the annotation process was carried out independently by two annotators with substantial domain knowledge, yielding a high inter-rater agreement (Cohen’s Kappa = 0.87). However, there is a potential for subtle subjective bias, particularly in handling ambiguous cases, such as distinguishing between privacy-related feature requests and bug reports. These edge cases often involve nuanced language, and consistent interpretation may not always be objectively accurate.

        Our dataset consists of user reviews from seven social media applications only, all sourced from the Google Play Store. This limits the generalizability of the GRACE model and the SENSOR tool across other app domains (e.g., finance, health) or app marketplaces (e.g., Apple App Store). Additionally, the data was restricted to English-language reviews, potentially reducing applicability in multilingual contexts.

    \subsection*{Construct Validity}
        We define three annotation classes --- \textit{privacy-related feature requests}, \textit{privacy-related bug reports}, and \textit{privacy-irrelevant} reviews. However, user reviews often contain overlapping themes or vague expressions that could be categorized under multiple headings. The rigid class boundaries may not always reflect the nuanced nature of user concerns, which might limit the construct validity of the model’s predictions. While GRACE performs well in distinguishing between predefined categories, future extensions of the tool could aim to detect emergent privacy themes via unsupervised clustering or zero-shot classification. Furthermore, our model evaluation primarily relies on traditional classification metrics (F1, ROC-AUC), which may not fully capture the practical utility or real-world impact of the tool in developer workflows.

    \subsection*{Conclusion Validity}
        While GRACE significantly outperforms baseline models on the given dataset, our conclusions regarding its superiority are outlined without testing for any statistical significance or feedback from user studies. Moreover, the deployment readiness of the model is inferred from computational metrics (e.g., inference time, model size), but has not yet been validated through real-world developer usage or feedback.

\section*{Discussion and Conclusion}   
    In this study, we underscored the practical effectiveness and scalability of SENSOR, a machine learning-enhanced online annotation tool designed specifically to assist developers in classifying privacy-related concerns in user reviews of social media applications. Unlike prior works that broadly categorized reviews into binary labels of privacy-related or not, SENSOR offers a more nuanced, three-way classification into \textit{privacy-related feature requests}, \textit{privacy-related bugs}, or \textit{privacy-irrelevant} content. This granular perspective allows developers to prioritize privacy-critical user feedback better and implement appropriate technical or policy-level interventions.

    As a utility tool for developers, the integrated features of SENSOR, including dynamic scraping of reviews, dual-mode annotation (manual and automated), and real-time monitoring of annotation progress and quality, make it stand out from previous systems. Existing annotation platforms often lack privacy-centric review mining and rarely provide role-based or automated annotation workflows. By bridging this gap, SENSOR supports both scalability and collaboration, empowering developers to address user privacy concerns.
    
    The performance of the proposed GRACE framework demonstrates the viability of combining GRU-based architectures with CBOW embeddings and attention mechanisms for text classification in privacy contexts. Achieving a macro-averaged F1-score of 0.9434 and ROC-AUC of 0.9934, GRACE surpassed all classical and deep learning baselines, including transformer-based models like RoBERTa and DistilBERT. Importantly, with a reasonable model size (44.09 MB) and a mean inference time (115.03 ms), GRACE also balanced classification accuracy (95.10\%) with inference efficiency, making it suitable for real-time deployment within the SENSOR platform. Furthermore, the reviews annotated using the SENSOR platform can be used to iteratively train the GRACE framework, broadening its classification abilities.

    Our results also reaffirm prior findings in the literature that classical models, while interpretable and lightweight, often struggle with nuanced and imbalanced datasets like user reviews. Despite achieving high precision for the \textit{privacy-irrelevant} class, classical models exhibited difficulty in distinguishing between \textit{privacy-related feature requests} and \textit{privacy-related bugs} due to overlapping linguistic features. Although the use of a hierarchical SGD variant partially addressed this limitation, its performance was still outpaced by RNN variants and transformer-based models.
        
    Building on the capabilities of SENSOR, several avenues for future work can further enhance its utility for privacy-aware app development. One promising direction is the integration of topic modeling techniques (e.g., LDA, BERTopic) \cite{abdelrazek2023topic} to extract recurring privacy themes from the annotated reviews. This would allow developers to identify dominant concerns within each category, providing a higher-level view of user sentiment and priorities. Second, we aim to extend the functionality of SENSOR to automatically extract and rank feature requests and bug reports from the annotated reviews, assisting developers in prioritizing actionable privacy improvements. Third, we plan to integrate SENSOR with a feature to train the model iteratively based on annotated reviews directly through the platform, enhancing the model's capability to generalize while overcoming the nuances of privacy-related reviews. Fourth, evaluate SENSOR and the GRACE framework on apps from other domains and in multilingual settings. Finally, to assess the practical utility of SENSOR in supporting privacy-aware development workflows, we plan to analyze its acceptability among developers and annotators using the Technology Acceptance Model (TAM) \cite{kabir2023acceptability}. The findings from this analysis will play a significant role in guiding future iterations of the platform and helping promote its adoption in real-world development settings.
    
    In conclusion, the SENSOR tool, powered by GRACE, offers a timely and robust solution for automated analysis of user reviews with a focus on privacy. Its strong classification performance, efficient inference capabilities, and collaborative annotation pipeline position it as a valuable addition to the software engineering and privacy-preserving application development toolkit. 

\nolinenumbers

\bibliography{SENSORBib}

\end{document}